\documentclass[12pt,preprint]{aastex}
\usepackage{amsmath}
\usepackage{graphicx}
\usepackage{epsfig}

\usepackage{url}
\usepackage{indentfirst}
\begin{document}
\title{A Model for  AR~Scorpii: Emission from  relativistic electrons trapped by  closed magnetic field lines
  of magnetic white dwarfs}
\author{Takata, J.\altaffilmark{1},  Yang, H.\altaffilmark{1} \and Cheng, K.S.\altaffilmark{2}}
\email{takata@hust.edu.cn }
\email{hrspksc@hku.hk}
\altaffiltext{1}{School of Physics, Huazhong University of Science and Technology, Wuhan 430074, China}
\altaffiltext{2}{Department of Physics, The University of Hong Kong, Pokfulam Road, Hong Kong}
\begin{abstract}
  AR~Scorpii is an intermediate polar system  composed of a magnetic
  white dwarf (WD) and an M-type star, and shows non-thermal, pulsed,
  and highly linearly polarized emission.
  The radio/optical emission modulates with the WD's spin and show the double
  peak structure in the light curves.  In this paper, we discuss a possible scenario for the radiation mechanism of AR~Scorpii. The magnetic interaction on the surface of the companion star
  produces  an outflow from the companion star, the heating of the companion star surface,  and the acceleration of electrons to a relativistic energy. 
The accelerated electrons, whose typical Lorentz factor
  is $\sim 50-100$, from the companion star  move along the magnetic field lines  toward the WD surface. The electrons injected with the
  pitch angle of $\sin\theta_{p,0}>0.05$ are subject to the magnetic
  mirror effect and are trapped in the
  closed magnetic field line region.We find that  the emission from the first magnetic mirror points
  mainly contributes to the observed pulsed emission and the formation
    of the double-peak structure in the light curve.
    For the inclined rotator, the pulse peak in the calculated light curve shifts the position in the spin phase,
    and a Fourier analysis exhibits a beat frequency feature, which are  consistent with the optical/UV observations.
    The pulse profile also evolves with the orbital phase owing  to the effect of the viewing geometry.
    The model also interprets the global features of the observed spectral energy distribution  in radio to X-ray energy bands.
  We also discuss the curvature radiation and the inverse-Compton scattering process in the outer gap accelerator of the WD in AR Scorpii and discuss the possibility  of the detection by future  high-energy missions.

\end{abstract}

\section{Introduction}
White dwarf (hereafter WD) is the end point of the stellar evolution of a progenitor with 
a mass of $M\leq8M_{\odot}$,  and  for over 97\% of all stars.
There are WDs with a  strong magnetic field in the range of $B_s\sim 10^{3}-10^9$G, which has been determined by the polarization and/or Zeeman splitting, and the WD belonging to this
class is called  a {\it magnetic WD} (Ferrario et al. 2015). The first magnetic WD, Grw+$70^{\circ}$8247 with 
$B_s=100-300$MG, was reported in the 1970s (Kemp et al. 1970; Angel et l. 1985), 
and it  is located about ~13{\rm pc} (43 lt-yr)  away. The number of the 
magnetic WDs is rapidly increased by the Sloan Digital Sky Survey (SDSS; York et al.
2000; Kepler et al. 2013). The WD catalog
based on  the SDSS DR7 (c.f. Kleinman et al. 2013) 
includes $19713$ WDs, in which  $\sim  12831$ is the 
hydrogen atmosphere WD (called DAs). Among the DAs listed in SDSS-DR7, 
Kepler et al. (2013) measured  the Zeeman splitting from $\sim 4\%$ (521) of 
all  DAs observed and estimated    
the surface magnetic field at  $B_s=1\sim 700$MG. In the SDSS DR10, Kepler 
et al. (2015) discovered 9089 new WDs, including 206 magnetic WDs. 
  Although the volume density of magnetic WDs  is still unknown, these observations would suggest that $5-10$\% of the WDs are magnetic 
  WDs (Sion et al. 2014; Kepler et al 2015). The space density of the WDs  estimated as $5\times 10^{-3}{\rm pc^{-3}}$ by the
  direct measurement of local WDs (Holberg 2002 and references therein)
  suggests that a large number of the magnetic WDs are existing in the galaxy.

  A magnetic WD is observed as an isolated system or a binary system, and the highly magnetic WD is
  found in the binary system, which is known as 
cataclysmic variable (hereafter CV). The CVs usually accrete matter on the WDs 
 from the late-type  main-sequence companion through Roche lobe overflow, and  
the magnetic CVs comprose $\sim 20$\% of all CVs 
($\sim 1100$ CVs; Ritter \& Kolb 2010).  The magnetic CV is mainly 
 divided into two groups, that is, polar and intermediate polar system (IP). 
The WD of the polar has a surface magnetic field of $B_s>10$MG, which is 
sufficiently large to lock the two stars into synchronous rotation with the orbital period ($P_{obs}\sim 100-500$minutes, Ferrario et al. 2015).  
 The WD in the IP has a  smaller surface magnetic field 
$B_s\sim 0.1-10$MG than that in the polar. The magnetic field of WDs 
is not sufficient to lock the companion into the synchronous rotation, and the 
spin of the WD is observed with a shorter period than the orbital period.
Both polars  and IPs emit X-rays owing  to accretion
of the matter from the companion on the WD's surface.

Since the synchronous rotation of two stars in polar, there is no angular momentum 
of the accreting matter relative  to the WD, and the formation of the accretion disk is prevented.
For the IPs, the accretion disk can present, and the angular momentum 
 transfer by the disk matter will spin up the magnetic WD. 
The accretion process on the  compact object is discussed with  the so-called co-rotation radius, $r_{co}=(GM_{WD}P_{WD}^2/4\pi^2)^{1/3}$,  where the WD's spin frequency is  equal to the disk rotation frequency 
(Keplerian frequency), and the Alfven radius, $r_M=3\times 10^{10}\dot{M}_{16}^{-2/7}M_{WD,1}^{-1/4}\mu_{WD,33}^{4/7}$cm (Frank et al. 2002), 
 where the magnetic pressure of the WD is equal 
to the dynamic pressure of the accretion disk.  Here $M_{WD}$ is 
the mass of the WD,
 $M_{WD,1}=M_{WD}/M_{\odot}$, $\dot{M}_{16}$ is the accretion rate in units of
 $10^{16}{\rm g~s^{-1}}$ and $\mu_{WD,33}$ 
is the dipole moment of the WD in units of  $10^{33}{\rm G~cm^{3}}$. 
When $r_{M}<r_{co}$, the inner edge of the accretion disk can enter  
inside the co-rotation radius and the disk matter can accrete
onto  the compact object. Then, the condition of the accretion through the disk 
may be written down as $\dot{M}_{16}\ge P_{WD,h}^{-7/3}\mu_{WD,33}^2M_{WD,1}^{-49/24}$, where  $P_{WD,h}=P_{WD}/1{\rm hour}$. 

Some  WDs in the IPs are rapidly rotating with a period 
$P_{WD}<100$s (Ferrario et al. 2015, and reference therein): $P_{WD}=33.1$s
 for AE~Aquarii (AE~Aqr), 70.8s  for DQ~Her, and  67.6s for V455~And.
For those magnetic WDs, an accretion rate close to the  Eddington rate
$\dot{M}_{E}\sim 10^{20}{\rm g/s}$ is necessary to satisfy the condition ($r_M<r_c$)
 for  the formation of the accretion disk. 
However, the  time-averaged rate of the  mass transfer from the companion 
 is observed with $\dot{M}_{16}=1-100$ (Patterson 1994) for those systems,  
suggesting  that these systems  will not contain the disk,  and probably 
the matter from Roche lobe directly streams  toward the WD surface, or most of the matter from the companion 
is centrifugally ejected (propeller phase).  In fact, 
the Doppler tomography profile of the AE~Aqr 
(e.g. Wynn et al. 1997) indicates that the AE~Aqr 
is in the propeller phase (Ikhsanov et al. 2004).  Moreover, 
the pulse timing study of the AE~Aqr revealed that the WD is spinning down at a 
rate of $\dot{P}_{WD}\sim 5.64\times10^{-14}$ (de Jager et al. 1994),
 corresponding to a spin-down luminosity of
 $L_{sd}=I(2\pi)^2\dot{P}_{WD}/P_{WD}^3\sim 6\times 10^{33}I_{50}{\rm erg/s}$, where
 $I_{50}$ is the WD's moment of inertia in units of $10^{50}{\rm g~cm^2}$. Because of the low 
mass accretion rate with the high spin-down rate, the AE~Aqr has been considered 
as the first WD pulsar candidate that operates the mechanisms of the neutron star (NS) 
pulsar-like particle acceleration and the non-thermal radiation process in the magnetosphere. 

The mechanism of the  NS pulsar-like particle acceleration 
in the magnetosphere of  the magnetic WD has been discussed
 by several authors (Usov 1988; Ikhsanov  1998; Kashiyama et al. 2011). 
 In the model, an electron (or a positron) that  emerged from the polar cap region is accelerated by an
 electric field parallel to the open magnetic field lines. The electric potential difference 
across the open field lines is estimated by 
\begin{equation}
V_{a}=(2\pi)^2\frac{\mu_{WD}}{2c^2P^2_{WD}}\sim 6\times 10^{13}
\left(\frac{\mu_{WD}}{10^{34}{\rm Gcm^3}}\right)\left(\frac{P_{WD}}{33{\rm s}}\right)^{-2}
{\rm static Volt},
\end{equation}
which can accelerate the electron up to 
$\gamma_{max}\sim eV_{a}/m_ec^2\sim 10^8\mu_{WD,34}(P_{WD}/33{\rm s})^{-2}$. This relativistic electron emits the nonthermal photon via the curvature radiation process with a  characteristic energy of 
\begin{equation}
E_{\gamma}=\frac{3}{4\pi}\frac{hc\gamma^3}{R_c}
\sim 200{\rm MeV}\left(\frac{\gamma}{10^8}\right)^3
\left(\frac{P}{33{\rm s}}\right)^{-1}
\left(\frac{R_c}{\varpi_{lc}}\right)^{-1},
\end{equation}
where $R_c$ is the curvature radius of the magnetic field line and $\varpi_{lc}=cP_{WD}/2\pi$ 
is the light-cylinder radius. Kashiyama et al. (2010) discussed 
 the mechanism of the NS pulsar-like pair-creation process in the magnetic WD magnetosphere  
as the possible  source of the cosmic-ray electrons and positrons. Although very high energy emission (de Jager 1994) and nonthermal X-ray emission (Terada et al. 2008) from 
AE Aqr were reported, the observational  view on the  WD pulsar has  not been
firmly established.  In section~\ref{discuss}, we will discuss the NS pulsar-like high-energy emission process
of the magnetic WD.

New discovery of the pulsed radio/optical/UV emission from AR Scorpii (hereafter AR~Sco) sheds light on the nonthermal nature of the magnetic  WD  (Marsh et al. 2016).
AR~Sco is the IP with an orbital period of  $P_{o}\sim 3.56$hr, and it  is composed of
an M star ($M_2\sim 0.3M_{\odot}$ and $R_2\sim 0.3R_{\odot}$)  and a magnetic WD.
The interesting properties of the emission from AR~Sco
are the nonthermal,  pulsed, and highly linearly polarized emission.
The radio/optical/UV emission modulates periodically on a period of  
$P\sim 1.97$minutes, which is thought to be  the spin period of the magnetic WD. The double-peak structure of the pulse profile and the morphology of the linear
polarization (Buckely et al. 2017) in the optical bands are
resemble to those of the Crab pulsar, which is the isolated young NS pulsar and emits the electromagnetic waves in radio to high-energy TeV bands (Kuiper et al. 2001; Kanbach et al. 2005; Takata et al. 2007). Moreover, the optical emission also modulates on the orbital period (3.56hr), which will reflect the heating of the 
dayside of the companion star by the interaction of the magnetic field/radiation 
by the WD. The modulation of the optical emission from the companion 
star with the orbital motion is also similar to that of the millisecond
NS pulsar/low-mass star binary system (Fruchter et al. 1988; Kong et al. 2012).  

AR~Sco's broadband electromagnetic spectrum in radio to X-ray bands is characterized by the synchrotron radiation from the
relativistic electrons, indicating
the acceleration process in the magnetosphere of the WD. As pointed out by Geng et al. (2016), on the other hand, the number of
the particles that emit the observed
nonthermal optical emission of AR~Sco is significantly larger than the number that can be supplied by the WD itself. This
suggests that the synchrotron-emitting electrons are supplied from the companion star's surface, and the acceleration process is different from that of the NS pulsar. Geng et al. (2016) suggested an electron/position beam from
the WD's polar cap sweeping the stellar wind from the companion star, and a bow shock propagating into stellar wind accelerates the electrons in the wind.

Geng et al. (2016) discussed the emission due to the interaction between the companion star
and the WD's open magnetic
field lines that extend beyond the light cylinder ($\varpi_{lc}=5.6\times 10^{11}$cm),  and hence
assumed that the WD is a nearly perpendicular rotator. In this paper,
we investigate another possibility that the electrons accelerated around the
companion stellar surface are trapped by the close magnetic field lines of
the WD.  The injected electrons  from the stellar surface are accelerated at the vicinity  of the companion star and  initially
travel toward the WD's surface along the magnetic field line.
We will solve  the evolution
of the pitch angle of such electrons  under the effects of the
synchrotron radiation energy-loss and the first adiabatic invariance. In section~2, we will describe our model and the basic equation for the motion
of the trapped electrons. We also discuss
the direction of the emission by the
relativistic electrons to calculate the light curve.  In section~3, we will
show our results  and discuss the mirror effect of the electron's motion.
We al show the model  pulse profile in optical and X-ray bands.
In section~4, we discuss the particle acceleration and the high-energy emission
from the outer gap accelerator of the WD in AR~Sco and will calculate
the expected fluxes of the curvature radiation process and
the inverse-Compton scattering (hereafter IC) process. 

\section{Theoretical Model}
\subsection{Energy injection}
In this section, we will discuss the emission model for the observed radio/optical/X-ray emission from AR~Sco.  The WD's magnetic field lines sweep periodically across the surface of the companion star. The strength of the magnetic field of the WD at the surface of the companion star will be of the order of 
\begin{equation}
B_{WD}\sim 195\left(\frac{\mu_{WD}}{10^{35}{\rm G cm^3}}\right)\left(\frac{a}{8\cdot 10^{10}{\rm cm}}\right)^{-3}{\rm G}.
\end{equation}
The modulation of the optical emission with the orbital period ($\sim 3.56$hr) indicates that the spinning  of the secondary star is synchronized with the 
orbital motion. With such a rapidly spinning M star, the stellar dynamo
 process can generate a polar magnetic field of several kG (Reiners et al. 2009). A magnetic interaction between the WD and M star will cause the magnetic reconnection or ohmic dissipation, and the dissipated  magnetic energy will be used for
 (1) the heating  of the M star surface, (2) acceleration of the electrons, and (3) outflow  from the M star.  

The magnetic interaction will produce an azimuthal component 
($\delta B_{\phi}$) of the magnetic field  of the WD, and its pitch 
$\eta=\delta B_{\phi}/B$ may increase  at $\eta\rightarrow 1$ 
before the magnetic field becomes unstable against  the magnetic 
dissipation process. In this model, we estimate the power of the magnetic 
dissipation as (Lai 2012; Buckley et al. 2017 and references therein)
\begin{eqnarray}
L_{B}&=&\frac{\eta B^2}{8\pi}(4\pi R_2^3\delta)\Omega_{WD} \nonumber 
\sim 2.8\times 10^{32} {\rm erg/s} \nonumber \\
&\times &\left(\frac{\mu_{WD}}{10^{35}{\rm G~cm^3}}\right)^{2}
\eta \left(\frac{\delta }{0.01}\right)\left(\frac{R_2}{3\cdot 10^{10}{\rm cm}}\right)^3 \left(\frac{a}{8\cdot 10^{18}{\rm cm}}\right)^{-6}
\left(\frac{P_{WD}}{117{\rm s}}\right),
\label{injection}
\end{eqnarray} 
where $\Omega_{WD}=2\pi/P_{WD}$,  $R_2$ is the radius of the  M star, and $\delta\sim 0.01$ is the skin depth (see  Buckley et al. 2017).

The magnetic interaction on the companion surface  may eventually cause an ablation of 
the matter from the M star surface and  an acceleration of the electrons to the relativistic energy. The ablation of the companion star
 by the WD magnetic field could occur  if the binary is close enough so that 
an energy deposition on the envelope of the stellar surface is high. 
The ablation of the companion star by the deposition of the  electromagnetic
 energy is a common process for the millisecond NS  pulsar and low-mass 
companion star binary, which is  called as  black widow/redback pulsars (Roberts 2013). In the black widow/redback systems, it has been observed that most of 
the deposited energy on the stellar surface is converted into the  heating of 
the companion star surface and/or the nonthermal radiation process, and 
a tiny  fraction (0.1-1\%) of 
 it is used for the ablation of the matter from the companion star
 (van den Heuvel \& Paradijs 1988; Cheng 1989).
 Since the mass loss  driven by the irradiation from the compact star has not been well understood,
 we introduce the efficiency factor $\chi$ for converting the dissipated energy 
into the kinetic energy of the wind, 
\begin{equation}
\chi=\frac{\dot{M}v^2_{esc}/2}{L_{B}}. 
\label{chi}
\end{equation}
For the black widow/redback system, the efficiency factor has been estimated as $\chi\sim 0.01-0.001$.

In terms of the efficiency factor, we estimate the rate of the particles leaving from 
the companion star surface as 
\begin{eqnarray}
\dot{N}_p&\sim&\frac{\dot{M}}{m_p}= 
\frac{\chi L_{B}}{\frac{1}{2}m_pv_{esc}^2} \nonumber \\
&\sim &5\times 10^{40}\chi\left(\frac{L_{B}}{10^{32}{\rm erg/s}}\right)
\left(\frac{v_{esc}}{5\cdot 10^{7}{\rm cm/s}}\right)^{-2}{\rm /s},
\label{dotn}
\end{eqnarray}
where $v_{esc}=\sqrt{2GM_2/R_2}$ is the escape velocity,  with
$M_2(\sim 0.3M_{\odot})$ being the mass of the companion star. Because of the charge conservation, we assume  that  the number of the electrons that leave from
the stellar surface  and that are accelerated by the magnetic dissipation process is of
the order of  $\dot{N}_e=\dot{N}_p$. 

In this model, we assume that most of the dissipated magnetic
energy is  used  for the acceleration of the electrons and/or the heating of the companion surface.
Hence, the typical Lorentz factor of the accelerated 
electrons may be  estimated  to be  
\begin{eqnarray}
  \gamma_0 &\sim& \frac{L_B}{\dot{N_e}m_ec^2}=\frac{m_pv_{esc}^2}{2\chi m_ec^2 }\nonumber \\
  &\sim& 50\left(\frac{\chi}{5\cdot 10^{-5}}\right)^{-1}
\left(\frac{v_{esc}}{5\cdot 10^{7}{\rm cm/s}}\right),
\label{lorentz}
\end{eqnarray}
where $\chi\sim 10^{-5}$ will be  chosen to fit the observed luminosity and SED  of AR~Sco.

\subsection{Motion of  trapped electron }
In AR~Sco, the light cylinder of the WD is $\varpi_{lc}\sim 5.6\times 10^{11}$cm, which is larger than the separation ($a\sim 8\times 10^{10}$cm) between two stars. Hence, the companion star
will interact with the close magnetic field lines of the WD, unless the WD is a nearly perpendicular rotator. As we discuss above, the magnetic interaction between the closed magnetic field of the WD and
the companion star will produce the pitch  $\delta B_{\phi}/B\sim 1$ for
the WD and will cause the magnetic reconnection/dissipation, which accelerates the electrons. After
sweeping across the companion star surface, the magnetic field line
of the WD will remain to  be closed, and the injected electrons may be trapped at the closed magnetic
field lines by the magnetic mirror effect. Since the radius of the gyration motion  of an electron 
is much smaller than the size of the magnetosphere, $r_{gy}\sim 440{\rm cm}(\mu_{WD}/10^{35}{\rm G~cm^{3}})^{-1}(\gamma_0/50)$, we  ignore any drift 
motion crossing the magnetic field lines in the {\it co-rotating frame} of the  WD, that is, in the laboratory frame,
the  trapped electron is co-rotating with the WD by the $\vec{E}\times \vec{B}$ drift,  and
it moves only  along the magnetic field line.  

 We expect that the observed optical emission from AR~Sco is produced by the
electron with the typical Lorentz factor $\gamma_0\sim 50$ of equation~(\ref{lorentz}). With the typical Lorentz factor 
$\gamma_0\sim 50$, on the other hand,  the time scale of the synchrotron  loss around the companion star is estimated as 
\[
\tau_{syn}\sim 400{\rm s}\left(\frac{\mu_{WD}}{10^{35}{\rm G cm^3}}\right)^{-2}
\left(\frac{a}{8\cdot 10^{10}{\rm cm}}\right)^{6}\left(\frac{\gamma_0}{50}\right)^{-1}, 
\] 
which is longer than the crossing time scale of $a/c\sim 2.5{\rm s}$. This
indicates that the injected electrons do not lose their energy around the
companion star, and  they  migrate into the 
inner magnetosphere before the synchrotron energy-loss. 
For the electrons moving toward the WD surface,
 the increase in the perpendicular momentum due to the first adiabatic 
invariance competes with the decrease in it due to the synchrotron loss.  

 The evolution of the Lorentz factor and the pitch angle 
 along the magnetic field line under the synchrotron energy-loss and the first adiabatic invariance
 may be described by (Harding et al. 2005)
\begin{equation}
\frac{d\gamma}{dt}=-\frac{P_{\perp}^2}{t_{s}},
\label{gamma}
\end{equation}
and 
\begin{equation}
  \frac{d}{dt}\left(\frac{P_{\perp}^2}{B}\right)=-2\frac{B}{t_s\gamma}
  \left(\frac{P_{\perp}^2}{B}\right)^2,
\label{perp}
\end{equation}
where $B$ is the local magnetic field strength, $t_s=3m_e^3c^5/(2e^4B^2)$, 
and $P_{\perp}=\gamma \beta\sin \theta_p$ with $\beta=v/c$ and $\theta_p$  the 
pitch angle. To solve above equations, we apply the pure dipole magnetic field of 
\begin{equation}
\vec{B}=\frac{3\vec{n}_p(\vec{n}_p\cdot\vec{\mu}_{WD})-\vec{\mu}_{WD}}
{r^3},
\end{equation} 
where $\vec{n}_p$ is the unit vector of the position. In this paper, $\alpha$ denotes the angle between the spin axis and the magnetic axis of the WD.

When  the synchrotron loss time scale is much longer than the crossing time scale, the perpendicular momentum of the electrons moving toward the WD surface from the companion star increases as 
\begin{equation}
P_{\perp}(r)=\left(\frac{a}{r}\right)^{3/2}P_{\perp,0},
\end{equation}
where $P_{\perp,0}$ is the initial perpendicular momentum. 
The magnetic mirror could  occur  at the point  
\begin{equation}
r_m\sim a\sin^{2/3}\theta_{p,0}
\end{equation}
where $\theta_{p,0}$ is the initial pitch angle, provided that the crossing 
time scale $\tau_{m,c}=r_m/c$ is shorter than  the synchrotron loss time scale 
at $r_m$, $\tau_{m, syn}=3m_e^3c^5a^6\sin^2\theta_{p,0}/(2e^4\mu_{WD}^2\gamma_0)$.  From the inequality  $\tau_{m,syn}>\tau_{m,c}$, the critical 
 initial pitch angle, above which
 the electron is subject to the magnetic mirroring, 
 may be written as 
\begin{eqnarray}
\sin\theta_{p,0}&>&\left(\frac{2e^4\mu_{WD}^2\gamma}{3m_e^3c^6a^5}\right)^{3/4} \\ 
\nonumber  
&\sim&0.03\left(\frac{\mu_{WD}}{10^{35}{\rm G cm^3}}\right)^{3/2}
\left(\frac{\gamma_0}{50}\right)^{3/4}
\left(\frac{a}{8\cdot 10^{10}{\rm cm}}\right)^{-15/4}.
\end{eqnarray}

\subsection{Radiation Process}
\label{radiation}
Besides the IR/optical/UV emission, the AR~Sco is also observed in the X-ray
bands, and this will indicate that the   electrons with a Lorentz factor
 larger  than $\gamma_0\sim 50$ exist in the magnetosphere of the WD.
To explain the X-ray emission,  we assume that a process related to 
 the magnetic dissipation on the companion star
 surface accelerates the electrons to the relativistic speed and  forms
 a power-law distribution of the electrons over  several decays in energy:
 \begin{equation}
f(\gamma)=K_0\gamma^{-p},~~\gamma_{min}\le\gamma\le\gamma_{max},
\label{distri}
\end{equation}
where we use $\gamma_{min}=\gamma_0$ of  equation~(\ref{lorentz}).
For the maximum Lorentz factor of the accelerated particle
is determined as the Lorentz factor at  which the synchrotron cooling timescale $\tau_s\sim 9m^3_ec^5/(4e^4B^2\gamma)$ is 
 equal to the acceleration timescale  $t_a\sim \gamma m_ec/(\xi eB)$,
 yielding $\gamma_{max}\sim 8\times 10^6 \xi^{1/2}(\mu_{WD}/10^{35}{\rm Gcm^3})^{-1/2} (a/8\cdot 10^{10}{\rm cm})^{1/2}$, where $\xi<1$ represents the efficiency of the acceleration.  By assuming the power-law index of $p\sim 2.5$, which is a fitting parameter, we calculate the normalization ($K_0$) and the 
  minimum Lorentz factor by solving  the conditions that
  $\int f(\gamma)d\gamma=\dot{N}_e$
  and $\int \gamma m_ec^2f(\gamma)d\gamma=L_{B}$.
 The spectrum of the synchrotron radiation at the photon energy $E_s$ 
is calculated from   
\begin{equation}
P_{syn}(E_s)=\frac{\sqrt{3}e^2B\sin\theta_p}{hm_ec^2}F_{sy}\left(\frac{E_s}{E_{syn}}\right),
\end{equation}
where $h$ is the Plank constant, $E_{syn}=3he\gamma^2 B\sin\theta_p/(4\pi m_ec)$ is the typical photon energy, and $F_{sy}(x)=x\int_0^{\infty}K_{5/3}(y)dy$ with $K_{5/3}$ being the modified Bessel function of the order $5/3$. 

\subsection{Radiation direction}
\begin{figure}[h]
  \centering
  \epsscale{1.0}
  \plotone{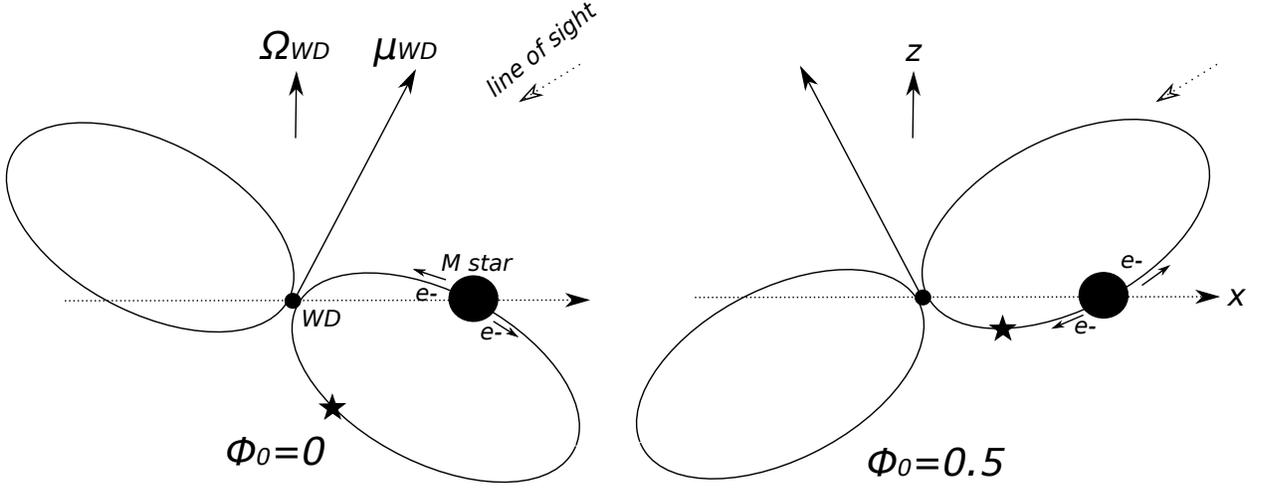}
  \caption{Schematic view of the AR~Sco system and the coordinate in the study. The observer is located
    within the plane made by the spin-axis (z-axis) and x-axis. The spin phase zero $\Phi_0$ is defined as when the magnetic axis points toward the observer. In the figure, the M star is assumed to be located
    at the plan by the spin axis and x-axis. The electrons are injected from the companion star by the
    magnetic field interaction. 
    The star symbols represent position of the first magnetic mirror point of the electrons injected
    into the southern hemisphere.  The travel time to the first magnetic mirror point is shorter
    for the electron injected at $\Phi_0=0.5$ than those injected at $\Phi_0=0$ (see section~\ref{pulse}). }
  \label{wd}
\end{figure}
To investigate the expected modulation of the observed emission  with the WD's spin, we calculate the propagation  direction and the time of arrival (TOA)
of each emitted photon. For the coordinate system, we introduce the $z$-axis at the spin axis of the WD and also assume that the spin axis of the WD and orbital axis are aligned for
 simplicity. The $x$-axis is chosen  so that the observer is located at the
 first quadrant in the ($x,~z$) coordinate (Figure~\ref{wd}).
 The direction of the magnetic momentum $\vec{\mu}_{WD}$ (the magnetic axis)
 is inclined by $\alpha$ from the $z$-axis, and it is rotating around the $z$-axis.

For a high Lorentz factor, we can anticipate that the emission direction of the particles coincides with the direction of the particle's velocity. In the laboratory frame, the 
unit vector of the electron motion that is co-rotating with the WD may be described by (Takata et al. 2007)
\begin{equation}
\vec{n}_e=\beta_0\cos\theta_p\vec{b}+\beta_0\sin\theta_p\vec{b}_{\perp}+\beta_{co}
\vec{e}_{\phi},
\label{emidi}
\end{equation}
where the first, second, and third terms on  the right-hand side represent 
the motion along the magnetic field line, the gyration motion, and the co-rotation motion, 
respectively. The value of the parallel speed $\beta_0$ in equation~(\ref{emidi})
at each point is calculated from the conditions that $|\vec{n}_e|=1$ and $\beta_{co}=\varpi/\varpi_{lc}$,
where $\varpi$ is the axial distance from the spin axis of the WD.
In addition, the vectors $\vec{b}=\vec{B}/B$, $\vec{b}_{\perp}$, 
 and $\vec{e}_{\phi}$ 
are the unit vectors along the magnetic field line, perpendicular to the magnetic field line, 
and in the azimuthal direction, respectively. The unit vector $b_{\perp}$ is defined by 
\begin{equation}
\vec{b}_{\perp}=\cos\delta\phi_g\vec{k}+\sin\delta\phi_g\vec{b}\times \vec{k},
\label{gyration}
\end{equation}
where $\vec{k}$ is  any unit vector perpendicular to the magnetic field line
and $\delta\phi_g$  refers to the phase of the gyration motion. 

The emission direction of equation~(\ref{emidi}) is interpreted 
as  the  angle measured from the rotation axis, $\zeta=\cos^{-1}n_{e,z}$,
where $n_{e,z}$ is the component of the emission direction  along the
rotation axis and the  azimuthal angle $\phi$ measured from the $x$-axis.
 Since the observer is laid on the first quadrant of the plane made by the $x$-axis
and $z$-axis (Figure~\ref{wd}),  we
 pick up the photon traveling in the direction  $\phi=0$ to calculate the pulse profile. 

The TOA  of the emitted photon measured on the Earth 
may be expressed by
\[
{\rm TOA}=t_{\Phi_{0}}+\delta t_{emi}+\frac{D-\vec{r}\cdot\vec{n}_e}{c},
\] 
where $D$ is the distance to the source.  
The first, second,  and  third terms on the right-hand side represent 
the time of the injection of the relativistic electron 
into the WD magnetosphere, the travel time  of the electron to the emission 
point after the injection, and the flight time of the emitted photon 
from the emission point ($\vec{r})$ to the Earth, respectively. We can safely ignore the effect of the flight time due to the
orbital motion of the WD, since it will be  $\delta t \ll  a/c\sim 2{\rm s}$, which is much smaller than the
spin period. The TOA can be translated into  
the spin phase of WD as 
\begin{equation}
\Phi=\Phi_0+2\pi\frac{\delta t_{emi}}{P_{WD}}
-\frac{\vec{r}\cdot\vec{n}_e}{\varpi_{lc}},
\label{phase}
\end{equation}   
where $\Phi$ is the spin phase at the detection of photon and $\Phi_0$ is
the spin phase at the injection of electrons. In this paper, we define $\Phi_0=0$ as the time when
the WD's magnetic axis is oriented in the plane made by WD's  rotation axis and the observer (Figure~\ref{wd}).  We note that the second term on the
right-hand side does not appear if the high-energy electrons are continuously injected on the
same magnetic field line with time, such as the high-energy emission from
the NS pulsar (Takata et al 2007). Moreover, since the main emission region
is  located at the position $r\ll a=8\times 10^{10}$cm, the third term on the right-hand side, $\sim r/\varpi_{lc}\ll a/\varpi_{lc}\sim 0.14$, is negligible, while  it is important
to produce the sharp pulse in the light curve of the high-energy emission from
the NS pulsar.

\section{Results}
\begin{figure}
\centering
\epsscale{1.0}
\plotone{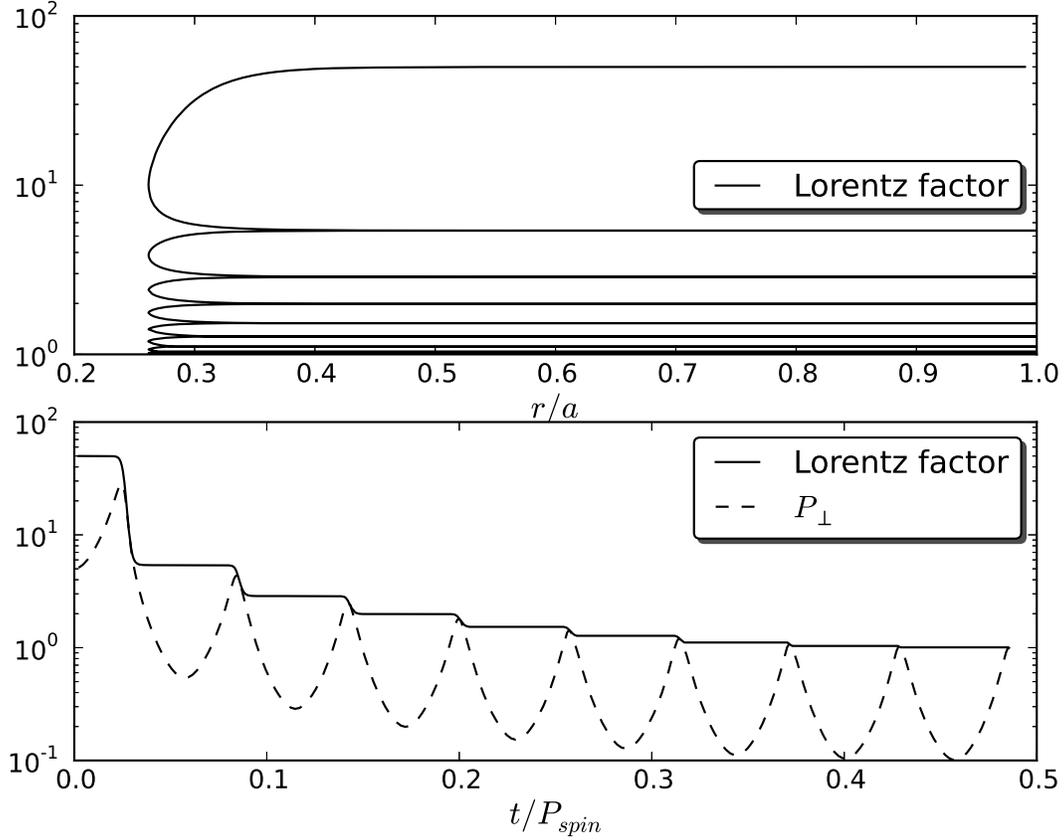}
\caption{Trajectory of the electron calculated from  equations (\ref{gamma}) and (\ref{perp}).
 Top panel: evolution of Lorentz factor along the radial distance. Bottom 
panel: evolution of Lorentz factor and perpendicular momentum ($P_{\perp}$) 
as a function of time. The magnetic field of the WD 
is assumed to be the pure dipole field with $\mu_{WD}=6.5\times 10^{34}{\rm G~cm^3}$,  and the magnetic axis is aligned with the spinning axis ($\alpha=0^{\circ}$). The electron with the pitch angle  $\sin\theta_{p,0}=0.1$ and the Lorentz factor $\gamma_0=50$ is injected 
from $r/a=1$ ($t/P_{WD}=0$) and the equator. 
The electron is trapped at the close magnetic field line
by the magnetic mirror.}
\label{traj}
 \end{figure}
\begin{figure}
\centering
\epsscale{1.0}
\plotone{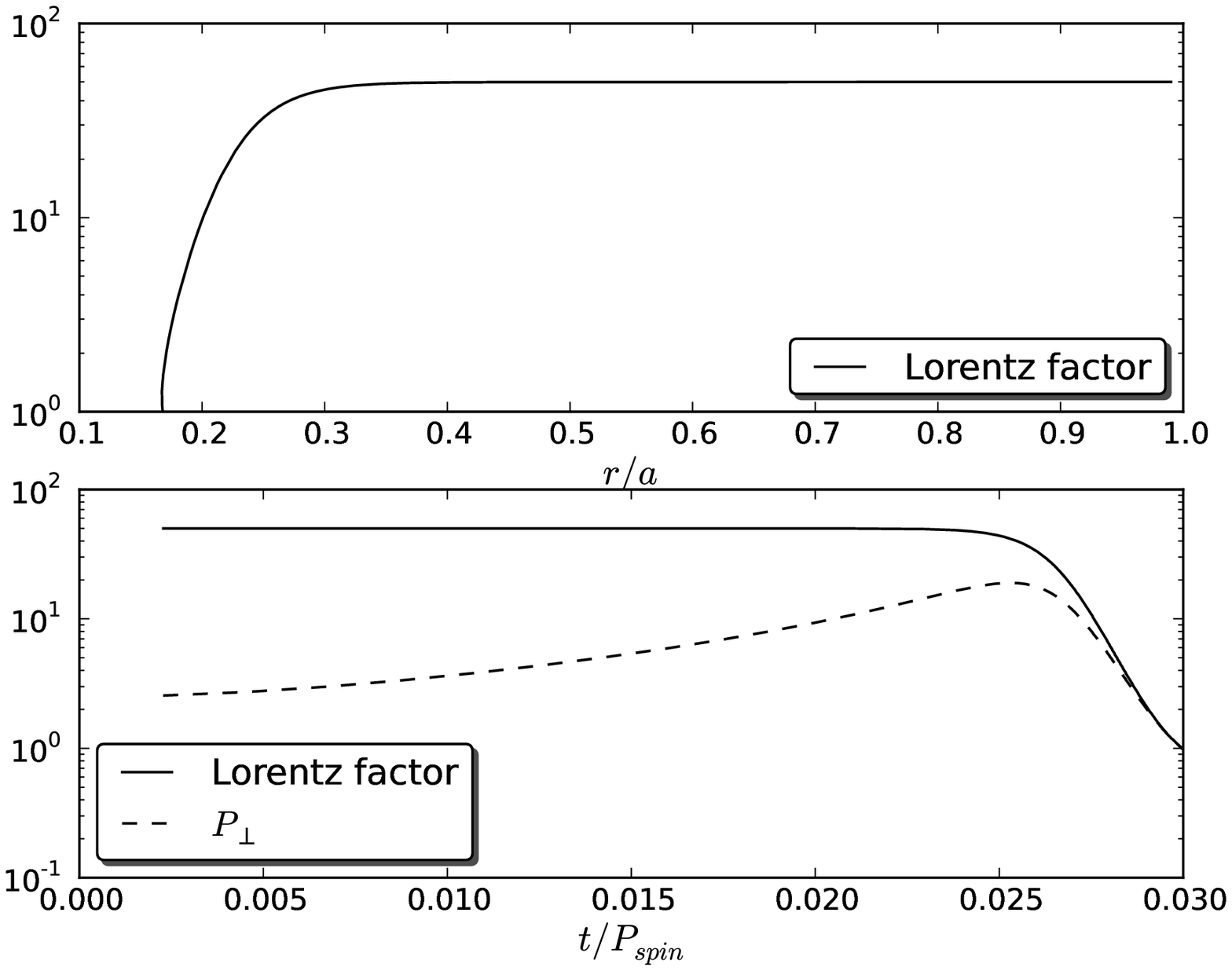}
\caption{Same as Figure~\ref{traj}, but $\sin\theta_{p,0}=0.05$. 
  The electron is not  trapped by the magnetic mirror. }
\label{traj005}
 \end{figure}
 
We treat the companion star as the point object located at 
$r=a=8\times 10^{10}$cm and on the equator of the WD, for simplicity. 
This  may be a rough treatment for this system,  
 since the size of the companion star is not negligible 
compared to the size of the orbital  separation.  However, we expect 
that the main results discussed in this paper will not be modified even if 
we take into account the size of the companion star. Under this assumption,  
we evaluate the magnetic energy dissipation and the 
injection of the electrons at the companion star. 

 \subsection{Magnetic mirror effect}
 \label{mirror}
Figures~\ref{traj} and~\ref{traj005} represent the evolution 
of the Lorentz factor and the perpendicular momentum ($P_{\perp}$) 
calculated from  equations~(\ref{gamma}) and~(\ref{perp}).
 In the calculations, the electron with the initial Lorentz factor $\gamma_0=50$ and 
the sine of the pitch angle $\sin\theta_0=0.1$ for Figure~\ref{traj} or $\sin\theta_0=0.05$ 
for  Figure~\ref{traj005} is injected toward the WD surface from $r=a=8\times 10^{10}$cm. We  assume
the magnetic dipole field with $\mu_{WD}=6.5\times 10^{34}{\rm G~cm^3}$ and the inclination angle of $\alpha=0^{\circ}$.

The top panel of Figure~\ref{traj} shows the evolution of the Lorentz factor as a function 
of the radial distance and indicates that the injected electron is trapped at
$\sim 0.25<r/a<1$ by the magnetic mirror effect. At around $r\sim a$, the synchrotron cooling time scale is longer than the dynamical time scale $\sim a/c$, and the electron migrates toward 
the WD with an almost constant Lorentz factor. As the dotted line in the bottom panel of Figure~\ref{traj} shows, 
the perpendicular momentum $P_{\perp}$ increases with a time owing
to  the first  adiabatic invariance. Around the magnetic mirror point, where the pitch angle becomes
$\sin\theta_p\sim 1$, the synchrotron energy-loss increases, and it rapidly decreases  the Lorentz factor.
After the magnetic mirror,
the electron moves toward the outer magnetosphere
 with the synchrotron loss time scale longer than the dynamical time scale,  
and hence the Lorentz factor is almost constant.
 Since the electron travels on the closed magnetic field lines, 
it eventually moves toward WD's surface  in another hemisphere and is reflected back again by the magnetic mirror. 
As Figure~\ref{traj} indicates, most of the initial energy
of the electron injected
 with  $\sin\theta_{p,0}=0.1$ is lost by the synchrotron radiation at the first magnetic mirror point.
 
\begin{figure}
\centering
\epsscale{1.0}
\plotone{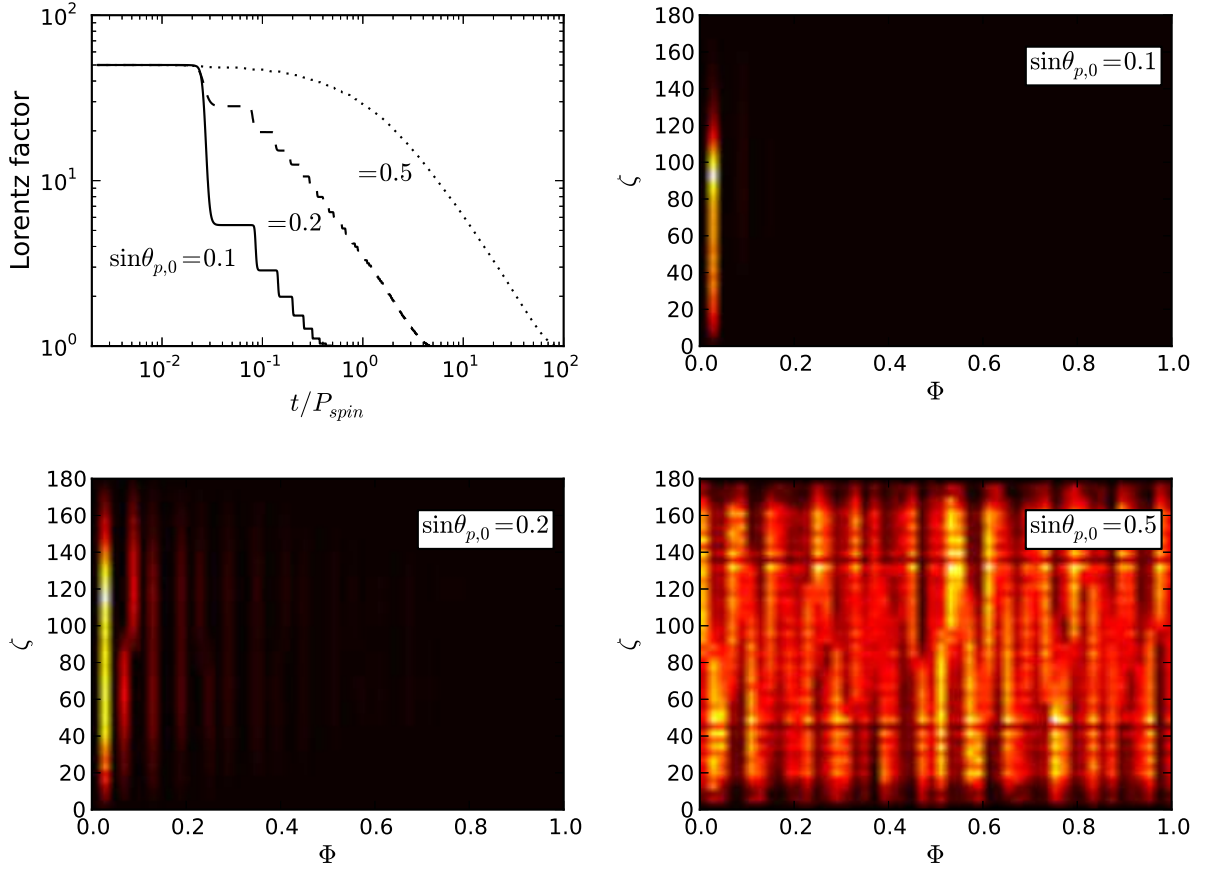}
\caption{Sky map of observing  angle ($\zeta$) and spin phase ($\Phi$) for the 
electron injected with the Lorentz factor $\gamma_0=50$ and 
 the pitch angle $\sin\theta_{p,0}=0.1$ for the top right panel, 
0.2  for the bottom left panel, and 0.5 for the bottom right panel. 
 The brightness of the color represents the intensity of the 
observed emission. The electrons are injected from $r=a$ and the equator at $\Phi_0=0$. 
The top left panel shows the evolution of the Lorentz factor as a function of time. The dipole magnetic  
field with  $\alpha=0^{\circ}$  is used for the calculation.
The effect of the gyration motion is taken into account in the calculation. }
\label{skymap}
 \end{figure}

For the smaller initial pitch angle of $\sin\theta_{p,0}=0.05$ in Figure~\ref{traj005},
the electrons can enter the  inner magnetosphere, and
the stronger synchrotron energy-loss prevents  the magnetic mirror. 
The top left panel of Figure~\ref{skymap} represents the evolution of the
Lorentz factor as a function 
of the time from the injection. The figure indicates that the electron injected 
with the  small pitch angle of $\sin\theta_{p,0}=0.1$  (solid line) 
radiates away all the initial energy within 
a  timescale less than the  spin period, while for the electron
with a larger pitch angle 
$\sin\theta_{p,0}=0.2$ (dashed line) and 0.5 (dotted line), the energy-loss time scale is longer 
than the spin period of the WD. 
\subsection{Formation of the pulse}
\label{pulse}
\begin{figure}[h]
  \centering
  \epsscale{1.0}
  \begin{tabular}{@{}cc@{}}
    \includegraphics[width=0.5\textwidth]{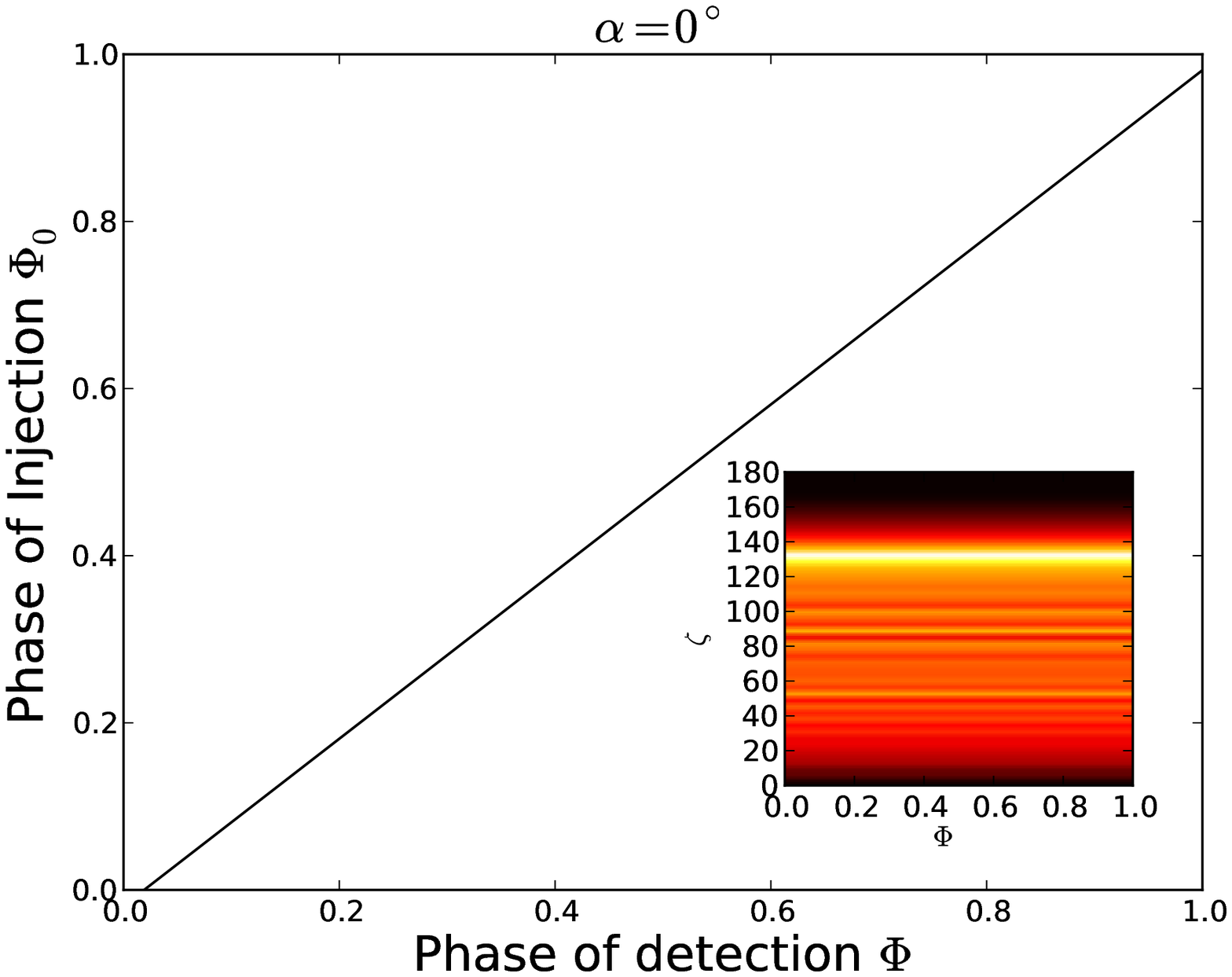} &
    \includegraphics[width=0.5\textwidth]{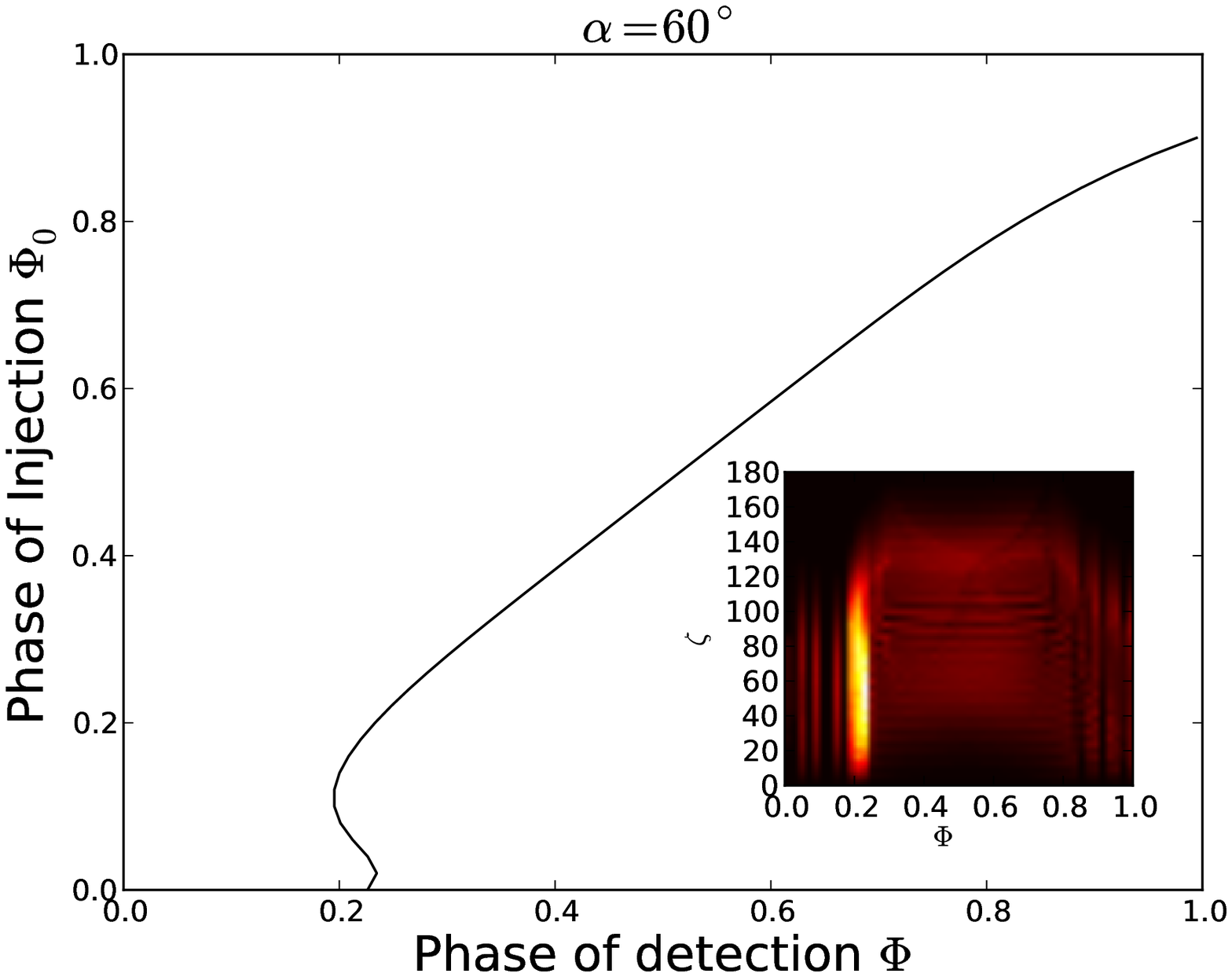} \\
  \end{tabular}
  \caption{Main panel: The relation between the phase of the injection
    of the electron ($\Phi_0$)
    and the spin phase of  the observed  emission from the first mirror point.
    The initial Lorentz factor
    and the pitch angle are given by $\gamma_0=50$ and $\sin\theta_{p,0}=0.1$, respectively.
    In addition, emission  from a specific gyration phase is calculated.
    Inset: sky map of $(\zeta,~\Phi)$. Emission from all gyration phases
    is included.}
  \label{phaser}
\end{figure}

Figure~\ref{skymap} shows  the sky map of the angle ($\zeta$) and the spin phase
 ($\Phi$)  for the emission from the electron injected at $\Phi_0=0$ with the Lorentz factor $\gamma_0=50$ and 
the  pitch angle  $\sin\theta_{p,0}=0.1$ (top right),  0.2 (bottom left), and 0.5 (bottom right); the brightness of the color in the figure refers the observed intensity. With $\gamma_0=50$,
the electrons injected with $\sin\theta_{p,0}\ge 0.1$ are subjected  to the magnetic mirror,  as the top left panel of  Figure~\ref{skymap} indicates, and the synchrotron  energy-loss mainly occurs at the magnetic mirror point. Around the magnetic mirror point, the pitch angle of the electron is increased to  90 degree, and the gyration motion produces  the synchrotron beam that
covers the large region  in the angle $\zeta$, as we can see in Figure~\ref{skymap}.

As we can also see in Figure~\ref{skymap}, the sky map of the observed emission from one electron depends on the initial pitch angle.  For the injected
electron with 
 $\sin\theta_{p,0}=0.1$ (see top right panel),
 since the most of the electron's energy  is radiated 
away  at the first magnetic mirror point, and the synchrotron emission  at 
subsequent magnetic mirror points is negligible. As a result, the emission from this electron can be mainly observed at narrow spin phase width.  For $\sin\theta_{p,0}=0.2$ (bottom left panel), 
  the magnetic mirror point is located outside of that for $\sin\theta_{p,0}=0.1$, and hence  the synchrotron energy-loss at the first magnetic mirror is less than that for $\sin\theta_{p,0}=0.1$. As we can
  see, however, the emissivity at the first magnetic mirror point is still much  higher than
the emissivity at subsequent mirror points. In our model, therefore, the emission at the
first mirror point provides the  most contribution on the observed emission from the AR~Sco.

For the larger pitch angle with $\sin\theta_{p,0}=0.5$ (bottom right panel), 
the electron is  trapped at $r\sim a$ and the time scale  of synchrotron
energy-loss is much longer than the spin period, as the dotted line in the
top left panel shows. The emission from this electron covers  almost the
whole sky,  and it may  be observed as a nonpulsed emission. Moreover,  such an
electron whose
cooling time scale is longer than the spin period will be absorbed by the companion star after one rotation of the WD and will not contribute much to the observed emission.

Since the WD is spinning,  the electrons will be injected on the different magnetic field lines
 at the different spin phases, that is,  $\Phi_0$ in equation (\ref{phase}).  We integrate the emission from all magnetic field lines  
to compare with the observations.  As we discussed above,
the emission from the first magnetic mirror point of the injected electron mainly
contributes to the observation.  In such a situation,  we find that the structure of the
pulse profile   mainly depends on how the spin phase of the detected photon  ($\Phi$ in equation (\ref{phase}))
is related to the injected  phase $\Phi_0$.  If the  phase of the detection
$\Phi$  monotonically shifts with the phase of injection $\Phi_0$,
 the predicted pulse profile does not significantly modulate with the spin period.  
 For example, the left panel in Figure~\ref{phaser} shows such a relation
 calculated with the magnetic inclination $\alpha=0^{\circ}$ and the  magnetic gyration phase $\delta\phi_g$ in equation~(\ref{gyration}); the small window in the figure shows the sky map of $\Phi$ and $\zeta$ with
 the contribution from the whole gyration phase.
 Because of the rotation  axisymmetric geometry
 of the magnetic field lines for  $\alpha=0^{\circ}$, the observed phase $\Phi$  monotonically 
shifts with the injected phase $\Phi_0$, as the figure shows. As a result, the observed intensity  for $\alpha=0^{\circ}$ is the constant with the time for any viewing  angle $\zeta$. 

For a large inclined rotator ($\alpha=60^{\circ}$), the phase $\Phi$ of the detected photon
from  the first mirror point does not monotonically shift with the phase of the injection
$\Phi_0$, as the line in  the right panel of Figure~\ref{phaser} indicates. We can
see in the figure that the photons emitted at   
the first mirror point by  the electrons injected at $0<\Phi_0<0.2$ are  observed at a narrower width in the spin phase around $\Phi\sim 0.2$. In the sky map of $\Phi-\zeta$, therefore, 
 the intensity at $\Phi\sim 0.2$ is  higher than the intensity at other phases, 
 as the inset in Figure~\ref{phaser} shows. As a result, the observer with
 an appropriate  viewing angle $\zeta$ will  measure a significant
 modulation of the emission with the spin period of the WD. 

The concentration of the detected photons at the narrower spin phase for the larger  inclination angle 
can be understood as follows. In the present calculation for Figure~\ref{phaser},
 we inject the electron into the  southern  hemisphere of the WD's magnetosphere (Figure~\ref{wd}). At 
 the companion star, the radial direction of the magnetic field of the WD
 depends on the spin phase.  For example, the radial direction at $\Phi_0=0$ (the magnetic axis points toward the companion) 
 points to the outer  magnetosphere, while  at $\Phi_0=0.5$ it is directed
 toward the inner  magnetosphere (Figure~\ref{wd}). For the electron 
injected at $\Phi_0\sim 0$, it initially moves toward the light cylinder and returns to the inner  magnetosphere along the closed magnetic field line, and therefore it takes a longer time, after
the injection,  to reach  the first magnetic mirror point than the electron that is injected at a  later time ($\Phi_0>0$).
This difference in the travel time to the first magnetic mirror point
causes an enhancement of the observed emission at the specific spin phase.
As we can see in the Figure~\ref{phaser}, the emission from the half-hemisphere makes 
one high-intensity region in the sky map. As we expect, therefore,
the contribution from the  opposite  hemisphere creates another high-intensity region, which
separates $\sim 0.5$ spin phase. The observer  whose line of sight cuts through two bright
regions in the sky map will measure the double-peak structure in the light curve. 

\subsection{Application to AR Scorpii}
\begin{figure*}
\centering
\plotone{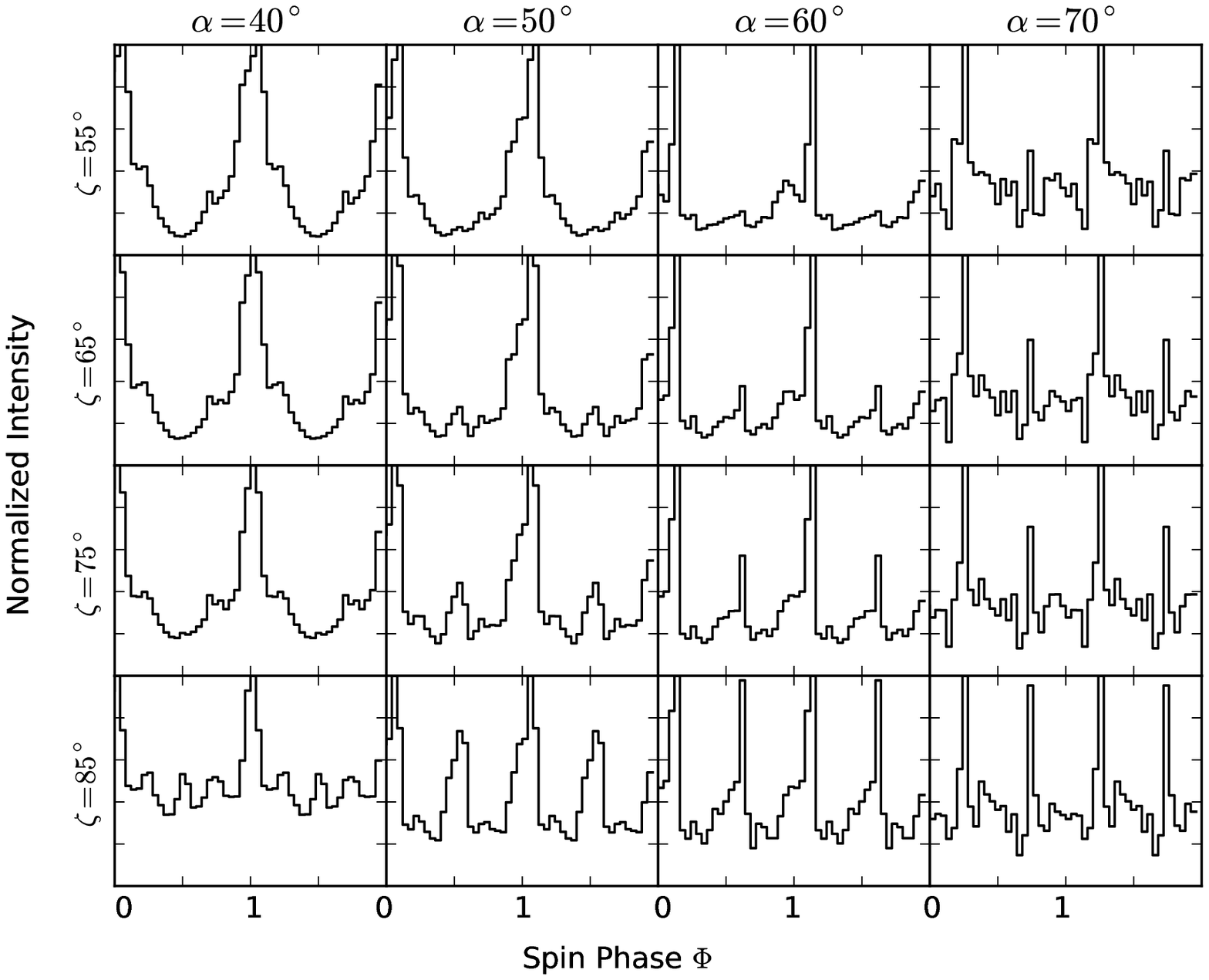}
\caption{Light curve in 0.1-1eV energy bands as a function of the inclination angle ($\alpha$) and the viewing angle ($\zeta$). The companion star is located
  between the WD and the  observer (inferior conjunction), and
  the phase zero in each panel
  corresponds to the time when the magnetic axis points toward the observer.
  The light curve of two  rotations of the WD is represented in each panel. }
\label{light}
 \end{figure*}
\begin{figure*}
\centering
\epsscale{1.0}
\plotone{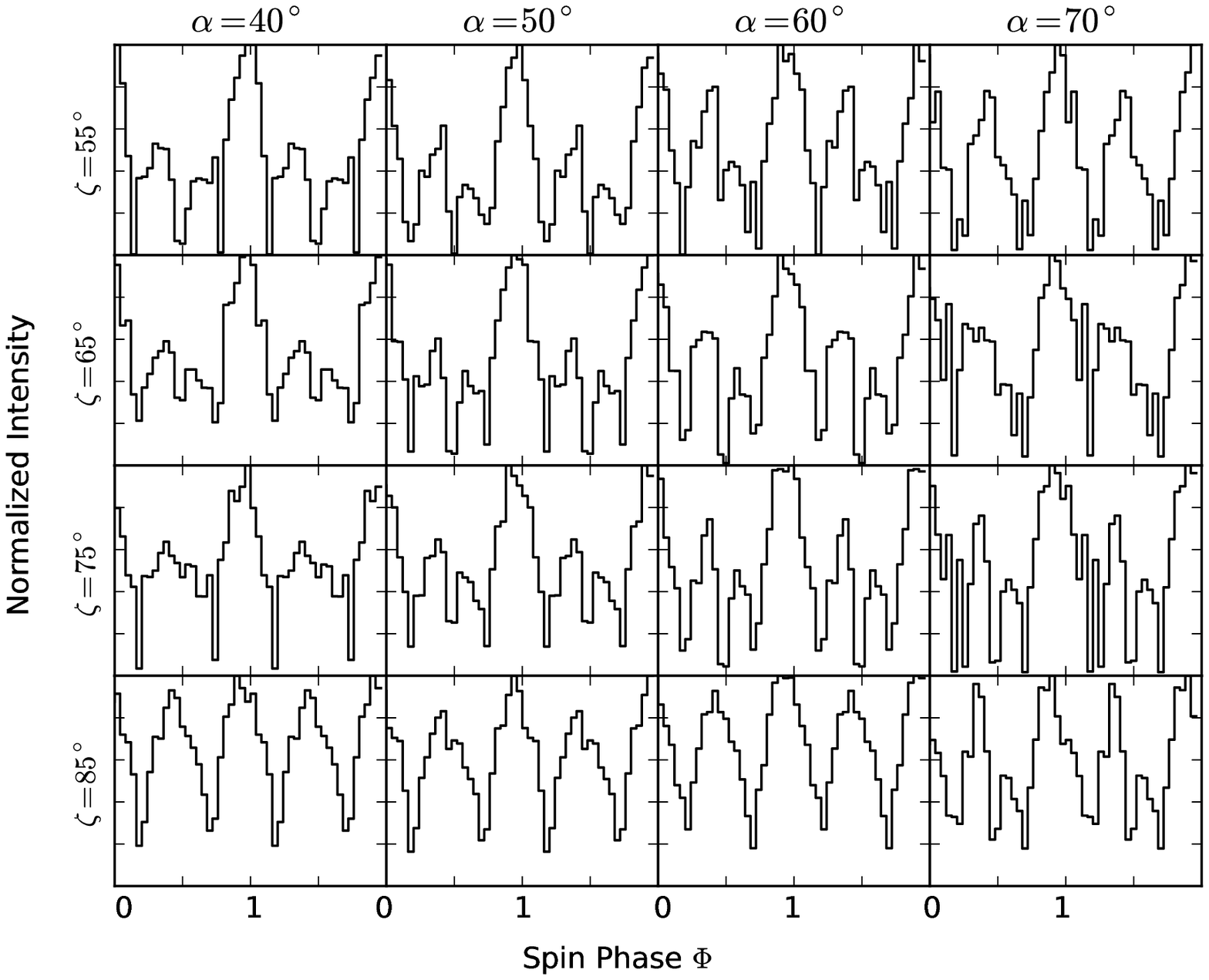}
\caption{Same as Figure~\ref{light}, but with 1-10keV energy bands.}
\label{lightx}
\end{figure*}
\subsubsection{Parameters and assumptions}
To apply the model to AR~Sco, we assume that  the magnetic interaction at the companion star injects the same number of the electrons  at the  northern
and southern hemispheres from the equator of the WD.
We represent the results with $\mu_{WD}=6.5\times 10^{34}{\rm G~cm^3}$, $\eta=1$, and  $\delta=0.01$ in equation~(\ref{injection})
and the efficiency $\chi=10^{-5}$ defined in equation (\ref{chi}).
 We fit the SED of the AR~Sco with $p=2.5$. 
With these parameters, the minimum and maximum Lorentz factors are estimated
as $\gamma_{min}\sim 50$ and $\gamma_{max}\sim 5\times 10^6$, respectively.
Other important parameters are the inclination angle $\alpha$ and
the viewing angle $\zeta$ that affect the predicted pulse profile. As we have assumed,
the spin axis of the WD is perpendicular to the orbital plane, and the WD is rotating in the same sense
as the binary orbit of the companion.

 As we discussed in section~\ref{mirror}, the evolution of the Lorentz factor and the perpendicular momentum of the injected electrons depend on the initial pitch angle. The  distribution
 of the pitch angle of the particles for different  acceleration processes
 has been
investigated by
the previous authors (Achterberg et al. 2001;  Kartavykh et al. 2016), but it is still
under investigation with a more realistic situation.
In this study, therefore, we assume an isotropic distribution in the
initial pitch angle $\theta_{p,0}$, that is,
\[
\frac{d\dot{N}_e}{d\theta_{p,0}}={\rm constant}. 
\]
We note that, in  our  model, the electrons that are
trapped at the close magnetic field line by the magnetic mirror are eventually 
absorbed by the companion stellar surface after one rotation   of the WD.
In the calculation, we do not take into account the synchrotron self-absorption, which may be important for
the radio band (Geng et al. 2016).

\subsubsection{Pulse profile}
Figures~\ref{light} and~\ref{lightx} summarize the predicted pulse profiles in 
optical (0.3-1eV) and X-ray (1keV-10keV) energy bands, respectively, at the inferior conjunction of the companion orbit, where the companion star is between the WD and the observer;
each panel  shows the pulse profile as a function of the magnetic inclination $\alpha$ 
and the Earth viewing angle $\zeta$. In addition, the each panel shows the 
pulse profile in the two spin phases of the WD  with normalized intensity, and the  phase zero
 (and unity) corresponds to the time when the magnetic axis points toward the
observer (Figure~\ref{wd}). 

We find that the model with a  smaller viewing angle predicts
a single pulse profile in the optical emission,
as Figure~\ref{light} indicates.
For the smaller inclination angle, the observer misses  the emission from
the half-hemisphere.
The optical/UV emission from AR~Sco has been observed as the pulsed emission
with the sharp double-peak structure in the light curve.
In the current model, the double-peak structure can be produced for  a larger magnetic inclination and a larger Earth viewing angle.
We note that  a larger viewing angle of the AR~Sco is  expected by the observed
orbital modulation of the optical emission from the companion star and by  a large amplitude of the modulation.

By comparing between Figures \ref{light} and~\ref{lightx}, we find that
 the  predicted X-ray pulse 
 is broader and the light-curve structure
 is more complicate in comparison with
  the optical light curve.  In the current model,
  the X-rays are produced by the Lorentz factor of
$\gamma>10^4$, for which the synchrotron cooling time scale is comparable to or shorter
than the crossing time scale. Unlike the optical emission, therefore,
the emission region is not restricted at the first magnetic mirror, but it spreads to
a wide space in the WD's magnetosphere.

\subsubsection{Orbital evolution of the pulse profile}
\begin{figure*}
  \centering
  \epsscale{1.0}
  \plotone{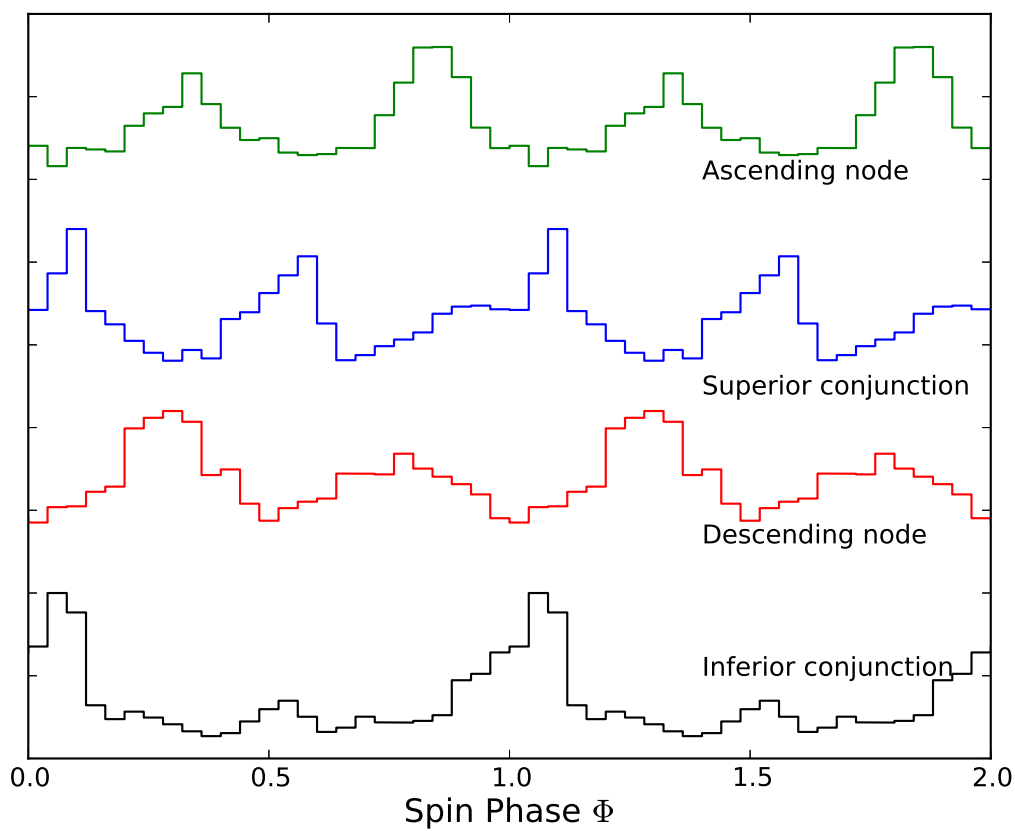}
  \caption{Orbital evolution of the pulse profile in the optical bands. The black,
    blue, red, and green histograms show the pulse profiles at the inferior conjunction,
    descending note, superior conjunction and ascending node, respectively, of
    the companion orbit. The phase zero corresponds to the time when the magnetic
    axis points toward the Earth. The inclination angle and the viewing angle
  are $\alpha=50^{\circ}$ and $\zeta=65^{\circ}$, respectively.}
  \label{lighto}
\end{figure*}

\begin{figure*}
  \centering
  \epsscale{1.0}
  \plotone{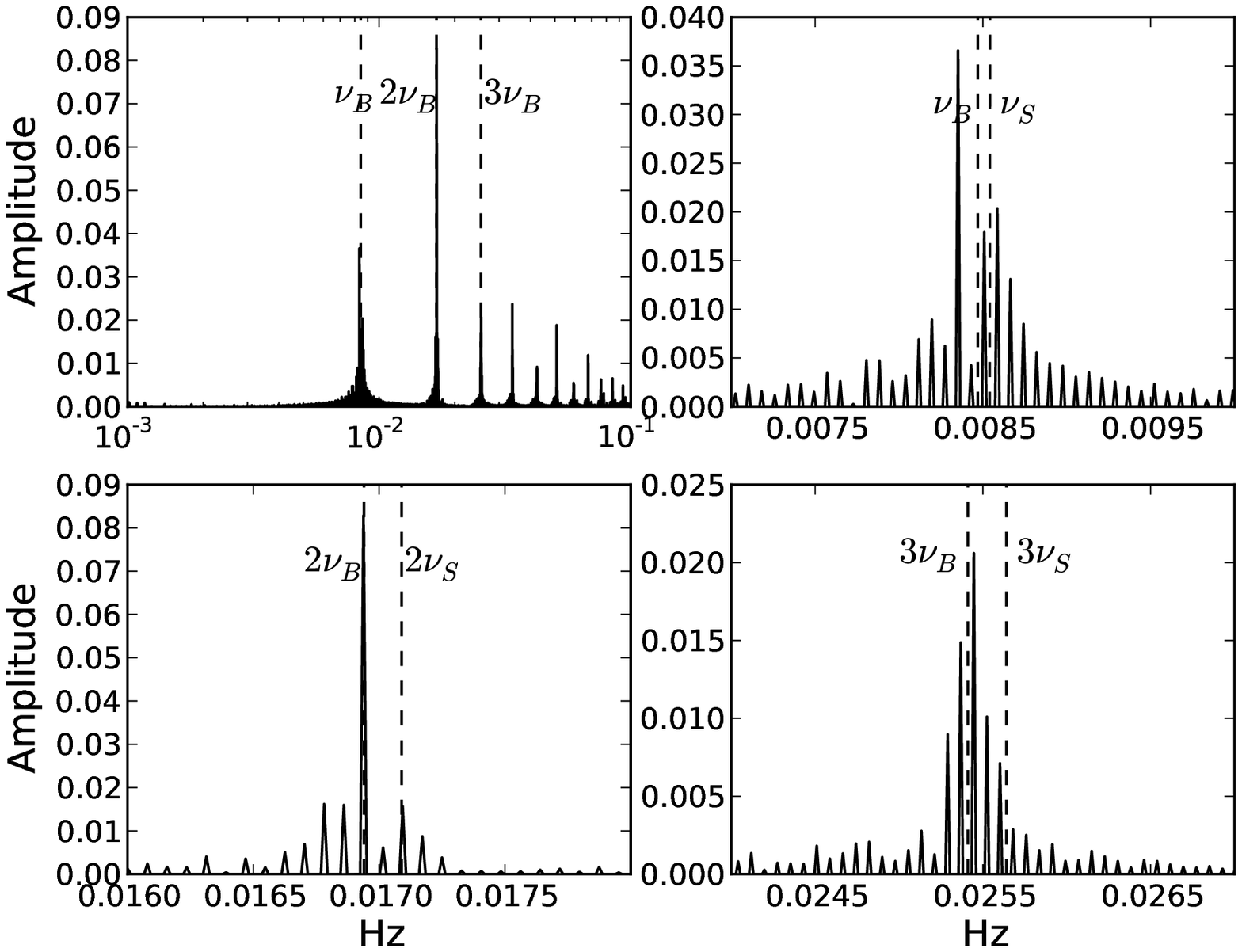}
  \caption{Power spectrum of the Fourier analysis for the model light curve calculated with the time of
    the three orbits. Tot left: global power spectrum. Top right: local power spectrum at the fundamental frequency.
    Bottom left: local spectrum at around the first harmonic frequency. Bottom right: local spectrum at around
    the second harmonic frequency. The locations for the fundamental, first harmonic, and second harmonic
  for the spin and beat frequencies are indicated with the dashed lines in the panels.}
  \label{powerspe}
  \end{figure*}
As we discussed in section~\ref{pulse}, the pulse peaks are mainly made
by the emission of  the electrons injected when the magnetic axis of the WD
is laid within the plane made by the WD's spin axis and the companion star. Because
of the orbital motion of the companion, therefore, the spin phase of the pulse shifts
with the orbital phase.  The pulse profile
presented in Figures~\ref{light} and~\ref{lightx} assumes the geometry
in which  the companion star is located at the inferior conjunction of the companion orbit, where the companion  is between the WD and the observer. Figure~\ref{lighto} compares the pulse profiles of the optical emission  at the inferior conjunction, descending node, superior conjunction, and ascending node, respectively, of the companion orbit. The phase zero in all panels corresponds to the time when the magnetic axis points toward the observer. We can see the shift of the pulse peaks and the evolution of the peak intensity with the orbital motion of the companion. We can also see the evolution
of the pulse profile along the orbit.

By assuming that  the WD is rotating in the same sense as the binary orbit of the companion, we make  the time sequence of the calculated intensity for about one orbit and perform the Fourier analysis of it.
Figure~\ref{powerspe} shows the power spectrum for the global frequency
(top left), around the fundamental frequency (top right), around
the second  harmonic frequency (bottom left), and around the third harmonic frequency (bottom right).
In the figure, we can see the peak of the power spectrum around the beat frequency $\nu_B=(1/117{\rm s}-1/3.56{\rm hours})\sim 0.0085$Hz and its harmonics,
reflecting that the pulse peak monotonically shifts along the orbit.  
In the figure, we also see that the first harmonic frequency has a stronger power, since the pulse profile has
a  double-peak structure.
These results are consistent with the power spectrum of the optical observations (Marsh et al.2016).

Marsh et al. (2016) reported no significant detection of the
pulsed emission in the X-ray bands. It could be due to the shift of the
pulse peak with the orbital phase, since the current X-ray instruments
cannot resolve the individual pulses of the AR~Sco. The time-averaged light
curve will be no pulsation or a modulation with a small pulse fraction. We may need a detailed analysis
for the timing of arrival of each X-ray photon to detect the pulsation.

\subsubsection{Spectrum}
\begin{figure}
  \centering
  \epsscale{1.0}
  \plotone{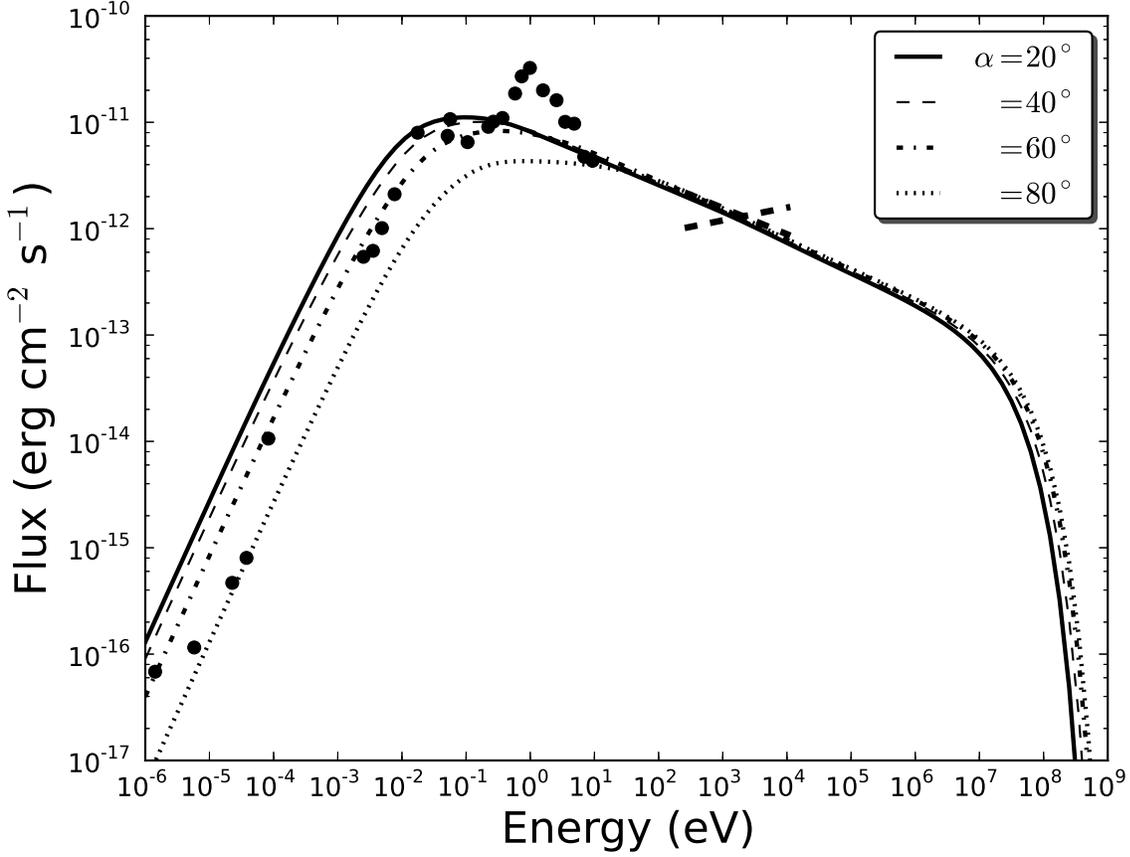}
  \caption{Spectrum of ARScorpii. The observational data
    (filled circles and thick dashed lines) without error bars are taken from
    Marsh et al. (2016). The different lines show the model calculations
    for the different inclination angles; $\alpha=20^{\circ}$ (solid), $40^{\circ}$ (thin-dashed),
    $60^{\circ}$ (dashed-dotted) and $80^{\circ}$ (dotted), respectively. In each model calculation,
   the emission from each electron is taken into account up to the time $t=P$ after the injection. 
   The results are for $\mu_{WD}=6.5\times 10^{34}{\rm G~cm^3}$, $\chi=10^{-5}$,
   $\gamma_{max}=5\times 10^{6}$ and $p=2.5$.}
  \label{spectrum}
\end{figure}

Finally, Figure~\ref{spectrum} compares  the observed   SED (Marsh et al. 2016) and model
calculations for the different inclination angles:
$\alpha=20^{\circ}$ (solid), $40^{\circ}$ (dashed),
$60^{\circ}$ (dot-dashed) and $80^{\circ}$ (dotted), respectively. In addition, we assume a power-law index of  $p=2.5$ of the injected electrons to
explain the broadband spectrum.  We take into
account the emission of each  injected electron up to $t=P$ (one rotation),
since the injected electron may be absorbed by
the companion star after one rotation of the WD.  We can see that the model interprets
the global features of the observed SED with the  spectral peak at $\sim 0.01$eV.
For $0.1-1$eV, the model flux is smaller than the observed flux. This is because the emission from
the companion star contributes to the observed flux in this energy band. In the figure, we can see that the different inclination angle predicts a slightly different model spectrum;
the SED of the larger  inclination angle has
a  spectral peak  at higher energy. This is because the magnitude of the magnetic field at the first magnetic mirror point tends to be
higher for larger inclination angle.  We note that the model SED
is insensitive to  the viewing  angle $\zeta$, because the radiation beam
from each electron at the mirror point covers a large region in the angle $\zeta$,
as Figure~\ref{skymap} shows.

\section{Discussion}
\label{discuss}
No concrete evidence  of the accretion onto the WD in AR~Sco has been found, and it is considered that
the system is now in  the propeller phase (Beskrovnaya \& Ikhsanov 2017). Hence,
if there is no flow of accretion matter from the companion star on the
open magnetic  field line region, the WD could  operate the mechanisms of the neutron star (NS)
pulsar-like particle acceleration and the  nonthermal radiation process in the magnetosphere. In this section,
we estimate the flux and typical photon energy of the emission  with a simple outer gap model.
The outer gap accelerator assumes the particle acceleration around the light cylinder (Cheng et al. 1986),
and it has been considered as the origin of the observed GeV gamma-ray emission from the NS pulsars (Abdo et al. 2013).

With the dipole magnetic field, the polar cap radius of the WD in AR~Sco is of the
order of
\[
R_{p}=R_{WD}\sqrt{\frac{R_{WD}}{\varpi_{lc}}}\sim 2.5\times 10^{7}\left(\frac{R_{WD}}{7\cdot 10^8{\rm cm}
}\right)^{3/2} {\rm cm}. 
\]
For the pulsar electrodynamics, the electric current emerges from the polar cap and is circulating in  the open field region.
The magnitude of the total electric
current will be given by the Goldreich-Julian value,
$I_{GJ}\sim \Omega_{WD} \mu_{WD}/2\varpi_{lc}$, (Goldreich \& Julian 1969).
The electric current running  through the outer gap is of the order of 
\[
I_{gap}\sim f_{gap} I_{GJ},
\]
where $f_{gap} (<1)$ is the fractional gap thickness, which is defined by
the ratio of the angular size of the gap thickness measured on the
stellar surface to the angular size of the polar cap $\theta_p=\sqrt{R_{WD}/\varpi_{lc}}$.

Using the fractional gap width $f_{gap}$, the available electric potential drop in the outer gap
is of the order of
\[
V_{gap}=f^2_{gap}V_{a}
\]
where $V_{a}=\mu_{WD}/(2\varpi^2_{lc})$ is the electric potential difference  between the magnetic axis and  the rim of the polar cap.  The current  carrier is accelerated in the
outer gap by the electric field parallel to the magnetic field
line. For the outer gap region,  the typical magnitude of the electric field can be estimated as 
\begin{equation}
E_{||}\sim \frac{V_{gap}}{\varpi_{lc}/2}\sim f_{gap}^2\frac{\mu_{WD}}{\varpi_{lc}^3}. 
\end{equation}
For the outer gap accelerator, the accelerated particles lose their energy via the curvature radiation process and the IC process.  The evolution
of the Lorentz factor ($\Gamma$) of an electron
along the magnetic field line   may be written as
\begin{equation}
  m_ec^2\frac{d\Gamma}{dt}=eE_{||}c-P_{curv}-P_{IC},
  \label{outtraj}
\end{equation}
where $P_{curv}$ and $P_{IC}$ are the energy-loss rate of the curvature
radiation and the IC process, respectively.
The loss rate of the curvature radiation is  given
by $P_{curv}=2e^2c\Gamma^4/(3R_c^2)$ with $R_c$ being the curvature radius of the
magnetic field line.  For an isotropic soft-photon field,
the energy-loss rate of
the IC process is calculated from 
\[
P_{IC}=\int\int(E_{\gamma}-E_s)\frac{\sigma_{IC}c}{E_s}\frac{dN_s}{dE_s}dE_sdE_{\gamma},
\]
where $E_{\gamma}$ and $E_{s}$ are the energies of the scattered and seed photons, respectively, and $dN_s/dE_s$ is the soft-photon field distribution. The cross
section is described by
\[
\sigma_{IC}=\frac{3\sigma_T}{4\Gamma^2}
\left[2q{\rm ln}q + (1+2q)(1-q)
  +\frac{(\Gamma_qq)^2(1-q)}{2(1+\Gamma_qq)}\right],
\]
where $\Gamma_q=4\Gamma E_s/m_ec^2$, $q=E_0/[\Gamma_q(1-E_0)]$ with
$E_0=E_{\gamma}/\Gamma m_ec^2$ and $1/(4\Gamma^2)<q<1$ (Blumenthal \& Gould 1970).

For the NS pulsar, the IC loss in the outer gap is negligible, and
the electric force is immediately  balanced with the radiation
drag force of the curvature radiation.
The saturated Lorentz factor is defined by equating between the acceleration term and the decelerating term
on  the right-hand side of  equation~(\ref{outtraj}),
\begin{equation}
  \Gamma_{sat}=\left(\frac{3R_c^2}{2e}E_{||}\right)^{1/4}
  \sim5\times 10^7f^{1/2}\left(\frac{\varpi_{lc}}{3\cdot 10^{8}{\rm cm}}\right)^{-1/4}
  \left(\frac{\mu_{NS}}{10^{30}{\rm G\cdot cm^3}}\right)^{1/4},
  \end{equation}
where $\mu_{NS}$ is the magnetic moment of the NS and we assume $R_c\sim\varpi_{lc}\sim 3\times 10^{8}$cm.
In the case of the saturated motion, the luminosity of the emission from the outer
gap is of the order of
\[
L_{\gamma}\sim I_{gap}\times \delta \Phi_{gap}\sim f^3L_{sd},
\]
where $L_{sd}$ is the spin-down power of the NS.
 
For AR~Sco, we can see that the IC energy-loss can be  comparable to or stronger
than the curvature radiation energy-loss. In the Thomson limit of the IC process, the energy-loss ratio
of the two radiation processes of AR~Sco becomes
\begin{equation}
  \frac{P_{IC}}{P_{curv}}=\frac{4\sigma_Tc\Gamma^2 U_{ph}/3}{2e^2c\Gamma^4/(3R_c^2)}\sim 1\left(\frac{R_c}{\varpi_{lc}}\right)^2\left(\frac{U_{ph}}{5\cdot 10^{-4}{\rm erg~cm^{-3}}}\right)\left(\frac{\Gamma}{3\cdot 10^7}\right)^{-2},
  \end{equation}
where $U_{ph}$ is the energy density of 0.01-1eV photons around the light cylinder
of the  WD.

The saturated motion of the Lorentz factor is  achieved    when
the maximum Lorentz factor accelerated by the electric potential drop $V_{gap}$ is
larger than the saturated Lorentz factor, that is, $eV_{gap}/m_ec^2>\Gamma_{sat}$.
For the AR~Sco, the saturation of the Lorentz factor will be achieved  when the fractional gap thickness satisfies the condition  that
\begin{equation}
  f_{gap}>1.4\left(\frac{\mu_{WD}}{10^{35}{\rm G~cm^3}}\right)^{-1/2}
  \left(\frac{\varpi_{lc}}{5.6\cdot 10^{11}{\rm cm} }\right)^{5/6},
\end{equation}
when the curvature radiation  is the  main energy-loss and
\begin{equation}
  f_{gap}>12\left(\frac{\mu_{WD}}{10^{35}{\rm G~cm^3}}\right)^{-1/2}
  \left(\frac{\varpi_{lc}}{5.6\cdot 10^{11}{\rm cm} }\right)^{1/2}
  \left(\frac{U_{ph}}{5\cdot 10^{-4}{\rm erg~cm^{-3}} }\right)^{1/2},
  \end{equation}
when the IC process is the main energy-loss. Since $f_{gap}$ should be less
than unity,   no saturation motion in the our gap accelerator will be realized for the WD in the AR~Sco. 

\begin{figure}
  \centering
  \epsscale{1.0}
  \plotone{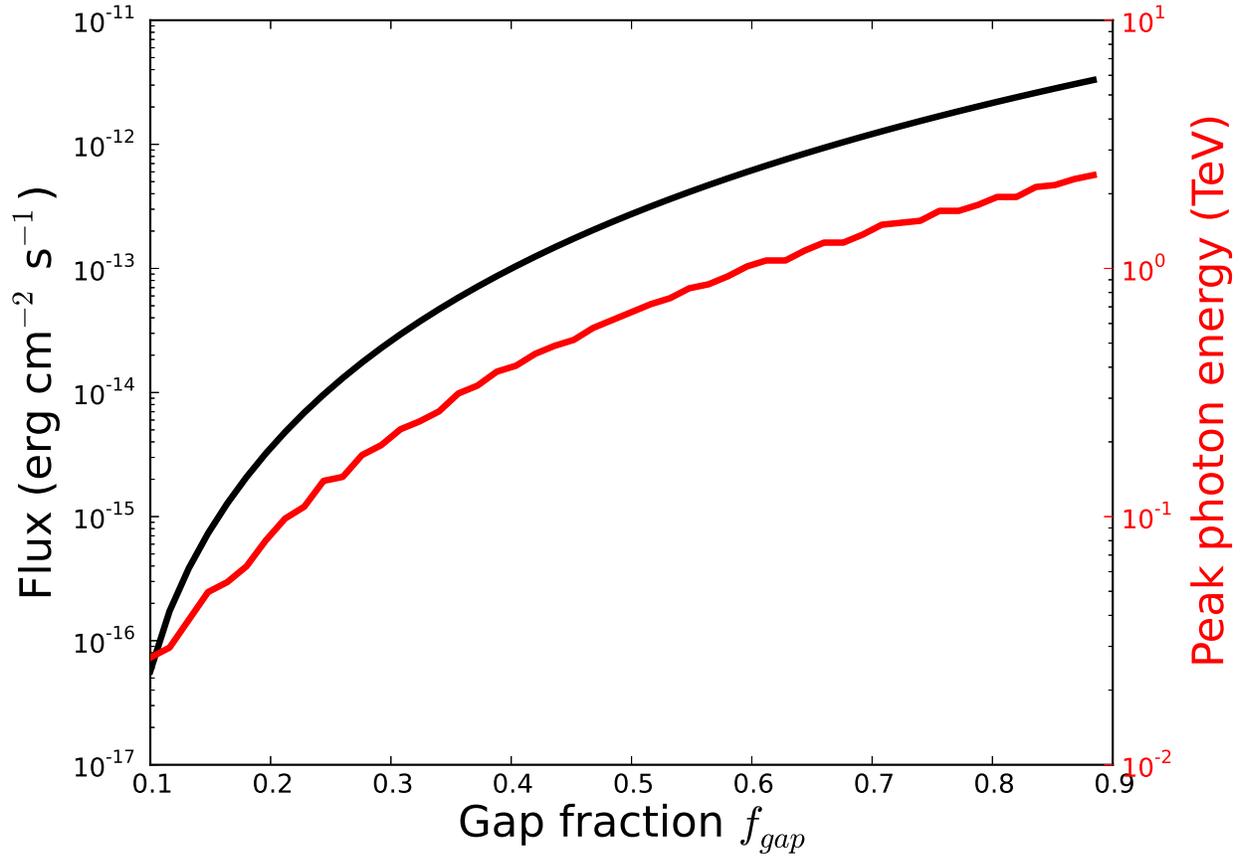}
  \caption{Predicted integrated flux (black) and peak photon energy in SED (red)
    of the IC process in the outer gap of AR~Sco's WD. The
    magnetic moment is assumed to be $\mu_{WD}=5\times 10^{34}{\rm G~cm^3}$. }
  \label{tev}
\end{figure}

For the NS pulsar, the fractional gap thickness will be determined by the pair-creation process between
the gamma rays from the curvature radiation process 
and the background X-ray from the stellar surface (Takata et al. 2012). For the WD pulsar, on other hand,
the fractional gap thickness could be determined by the photon-photon pair creation process of the TeV photon from
the IC process. The optical depth of the TeV photon is estimated as $\tau_{p}\sim \sigma_{\gamma\gamma}n_{opt}\varpi_{lc}
\sim 10^{-3}(n_{opt}/10^{10}{\rm cm^{-3}})$, where $\sigma_{\gamma\gamma}\sim \sigma_T/3$ is the cross section
of the pair-creation and $n_{opt}$ is the number density of the  $\sim 1$eV photon around the light cylinder. For the WD in AR~Sco,
the fractional gap thickness $f_{gap}\sim 1$ is required to produce TeV photons in the outer
gap (Figure~\ref{tev}).  If a part of the open field line region is occupied by the matter from the companion star, the possible
fractional gap thickness is  less than unity. Because of the theoretical uncertainties,
we parameterize the fractional gap thickness in this section.

\begin{figure*}
  \centering
  \epsscale{1.0}
  \begin{tabular}{@{}cc@{}}
    \includegraphics[width=0.5\textwidth]{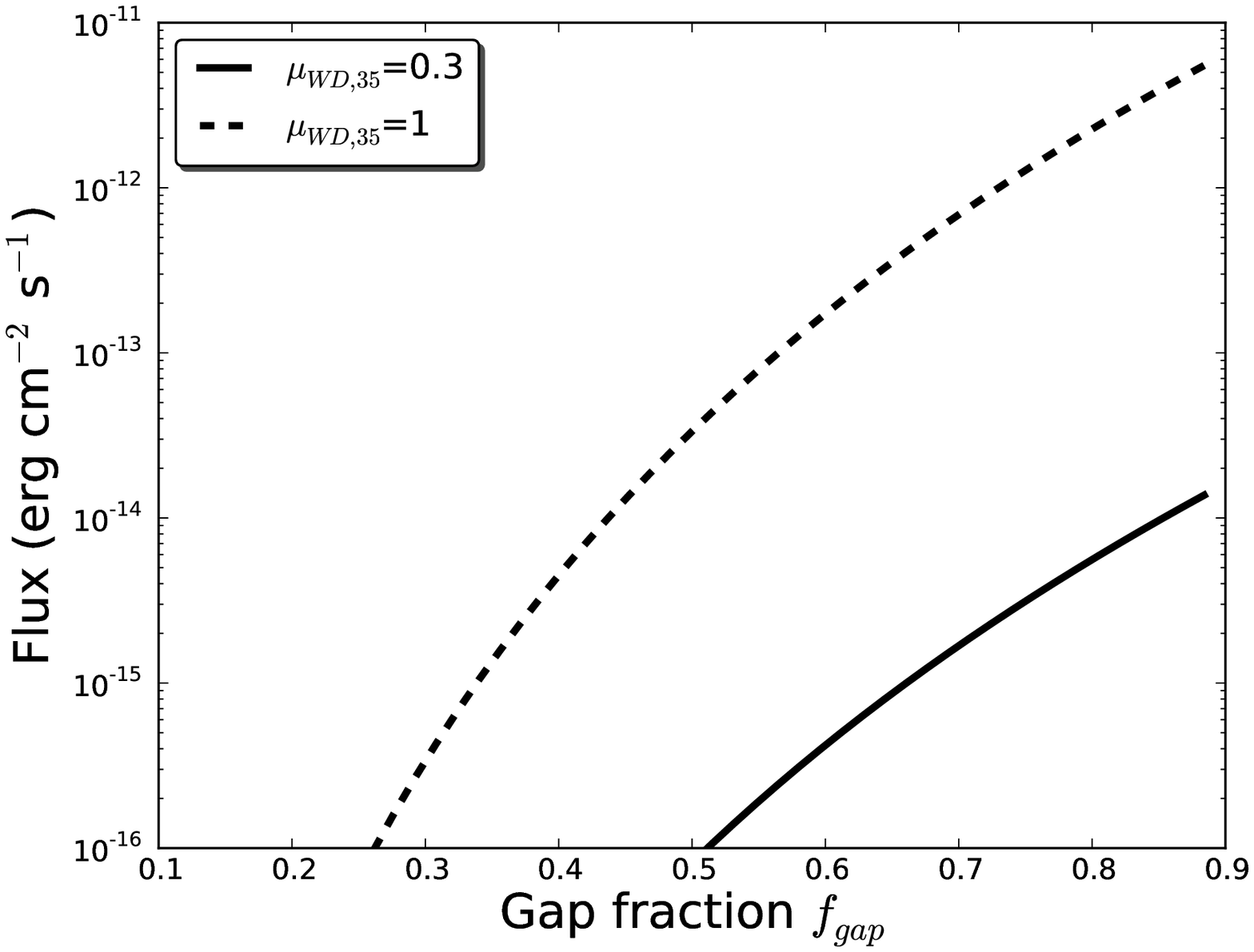} &
    \includegraphics[width=0.5\textwidth]{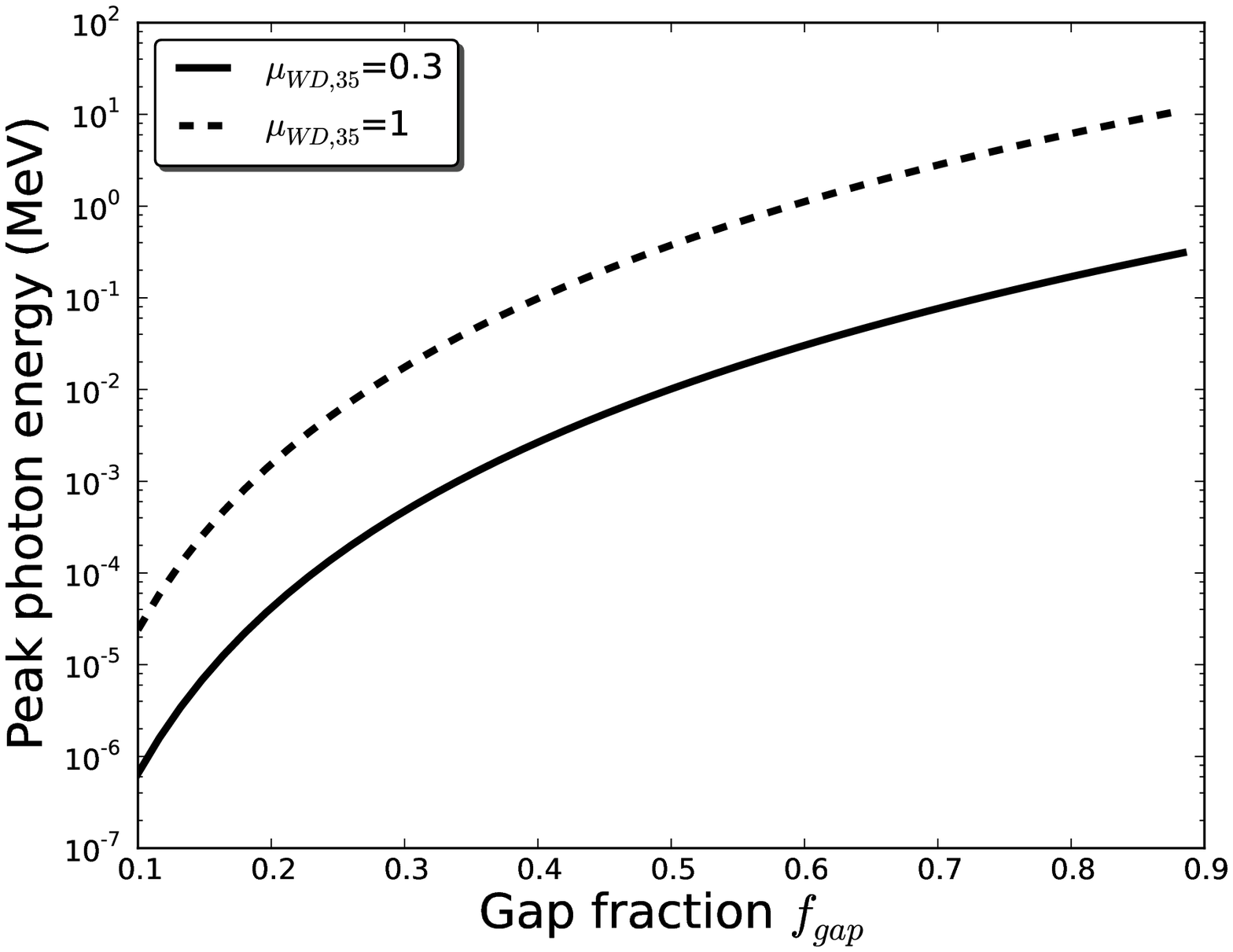} \\
  \end{tabular}
  \caption{Predicted integrated flux (left) and peak photon energy in SED (right)
    of the curvature radiation process in the outer gap of AR~Sco's WD. The
  dashed and solid lines are the magnetic moment of $\mu_{WD}=3\times 10^{34}{\rm G~cm^{3}}$ and $10^{35}{\rm G~cm^{-3}}$, respectively.}
  \label{outgap}
  \end{figure*}

Figure~\ref{outgap} summarizes the integrated flux (left panel) measured on the Earth
and the peak photon energy (right panel) of the curvature radiation process in the outer gap of AR~Sco
as a function of the gap fraction $f_{gap}$; in the figure, the solid and dashed lines show the results
for the magnetic moment of $\mu_{WD}=3\times 10^{34}{\rm G~cm^3}$ and $10^{35}{\rm G~cm^3}$, respectively, and
the solid angle of $4\pi$ rad is assumed. The figure indicates that if the gap fraction is larger than
$f_{gap}\sim 0.7$, the future hard X-ray/soft gamma-ray missions
would  measure the curvature  radiation from the AR~Sco. With a  magnetic moment $\mu_{WD}\sim 10^{35}{\rm G~cm^3}$, for example, the flux measured on the Earth will be  $10^{-12}~{\rm erg/cm^2}$ and the
peak photon energy of the spectrum is in $E=0.1-10$MeV for $f_{gap}>0.7$. This sensitivity could be achieved by
the future hard X-ray/soft gamma-ray missions.

The soft X-ray emission from AR~Sco is observed with a flux $F\sim 5\times 10^{-12}{\rm erg~cm^{-2}~s^{-1}}$ (Marsh et al. 2016), and it will not be the origin of  the outer gap. The current model predicts the gap fraction $f_{gap}\sim 0.3$
to produce the soft X-ray with the curvature radiation process. With $f_{gap}\sim 0.3$, however, the
predicted flux is much smaller than the observation. If the fractional gap width is $f_{gap}<0.2$, the peak
photon energy of the outer gap emission is located in the optical band. In such a case, however, the outer gap
emission is buried under the emission from the trapped electrons/stellar emission discussed in the previous
sections.

Figure~\ref{tev} shows the predicted flux measured on the Earth and the peak photon
energy of the IC process of the outer gap. As we can see in Figure~\ref{tev},
the IC process produces $>0.1$TeV photons if the gap fraction is $f_{gap}>0.3$.
Moreover, for $f_{gap}>0.5$,
the expected flux is $>10^{-13}{\rm erg~cm^{-2}~s^{-1}}$, which would be measurable
by the future CTA observation.

In summary, we discussed the radio/optical/X-ray emission from AR~Sco. In our model,
the magnetic dissipation process  on the M star surface produces
(i) an outflow from the companion star, (ii) heating of the
companion star surface, and (iii) acceleration of the electron to relativistic energy.  The accelerated electrons, whose typical Lorentz factor is
$\gamma_0\sim 50$, are trapped in the close
magnetic field lines. We solved the motion of the trapped electrons along  the magnetic field under the effects of
the synchrotron loss and the first adiabatic invariance. We found 
that the electron injected toward the WD's surface
with a pitch angle of $\sin\theta_{p,0}\ge 0.05$ is trapped at the
closed  magnetic field lines by the magnetic mirror effect.  For such an electron,
most of the  initial energy is released at the first mirror point by the synchrotron  radiation process. We
demonstrated that for the inclined rotator with the dipole
magnetic field, the synchrotron emission from the trapped electrons injected at the different spin phase
can  create the light curve with a double-peak structure, which is consistent with the observations.
The model expects that the pulse profile is linearly shifts with the orbital phase, and this shift is observed as
the beat frequency in the spectral power of the Fourier analysis. The model interprets
the global features of the observed SED in radio to X-ray energy bands. We discussed
the curvature radiation and the IC process in the outer gap
accelerator of the WD in AR~Sco.  The curvature radiation from the outer gap could be
measured by the future hard X-ray and soft gamma-ray missions, if the gap fraction
is $f_{gap}>0.7$. The TeV emission via the IC process from AR~Sco
may be also detected  by the future CTA project. 

We express our appreciation to an anonymous referee
for useful comments and  suggestions.
We thank to Drs. Lin, L.Chen-Che and Hu, Chin-Ping for useful discussion on the timing analysis.
J.T. and H.Y. are supported by NSFC grants of Chinese Government under 11573010, U1631103, and 11661161010. K.S.C. is supported by GRF grant under 17302315.

\end{document}